\documentclass[useAMS,usenatbib]{mn2e}

\usepackage{amssymb}
\usepackage{amsfonts}
\usepackage{mncite}
\usepackage{epsfig}
\usepackage{psfig}

\newcommand{\de}{\mbox{d}}

\newcommand{\msunyr}{M_{\sun}/\mbox{yr}}
\newcommand{\mdot}{\dot{M}}
\newcommand{\uno}{({\it i}) }
\newcommand{\due}{({\it ii}) }
\newcommand{\tre}{({\it iii}) }

\defcitealias{bellin94}{BL94}

\begin{document}

\title[Massive planets in FU Orionis]{Massive
  planets in FU Orionis discs: implications for thermal instability models}

\author[G. Lodato \& C. J. Clarke]{G. Lodato and C. J. Clarke\\
Institute of Astronomy, Madingley Road, Cambridge, CB3 0HA\\}

\maketitle
\begin{abstract}
FU Orionis are young stellar objects undergoing episodes of enhanced
luminosity, which are generally ascribed to a sudden increase of mass
accretion rate in the surrounding protostellar disc. Models invoking a
thermal instability in the disc are able to reproduce many features of
the outburst, but cannot explain the rapid rise time-scale observed in
many cases. Here we explore the possibility (originally suggested by
\citealt{clarkesyer96}) that the thermal instability is triggered away
from the disc inner edge (at a distance of $\approx 10R_{\odot}$ from
the central protostar) due to the presence of a massive planet
embedded in the disc. We have constructed simple, one-dimensional,
time-dependent models of the disc evolution, taking into account both
the interaction between the disc and the planet, and the thermal
evolution of the disc. We are indeed able to reproduce rapid rise
outbursts (with rise time-scale $\approx 1$ yr), with a planet mass
$M_{\mathrm{s}}=10-15M_{\mathrm{Jupiter}}$. We show that the luminosity
and the duration of the outbursts are increasing functions of planet
mass. We also show that the inward migration of the planet is
significantly slowed once it reaches the radius where it is able to
trigger FU Orionis outbursts, thus suggesting that a single planet may
be involved in triggering several outbursts.
\end{abstract}

\section{Introduction}

FU Orionis objects are a small class of Young Stellar Objects
undergoing violent and probably recurrent outbursts, during which they
can increase their bolometric luminosity by two to three orders of
magnitude \citep{hartmann96}. These outbursts are generally attributed
to a sudden increase of the mass accretion rate (which can reach
$10^{-4}M_{\odot}/$yr) in the disc of an otherwise ``normal'' T Tauri
star. This interpretation is suggested by a number of observations:
({\it i}) in one case (i.e. V1057 Cyg, \citealt{herbig77}) a
pre-outburst stellar spectrum is available, and it shows typical
features of a T Tauri star; ({\it ii}) the Spectral Energy
Distribution (SED) after the outburst is well described in terms of
accretion disc SEDs \citep{kenyon91,LB2001} ({\it iii}) optical and
near infrared line profiles are usually double-peaked, as expected if
they are produced in a differentially rotating disc
(\citealt{kenyon88,LB03b}; see however \citealt{herbig03} for a
dissenting view).

The light curves of the three best studied FU Orionis objects (i.e.,
FU Ori itself, V1515 Cyg, and V1057 Cyg) show remarkable differences
between each other. The rise timescale of FU Ori and of V1057 Cyg is
very short (of the order of 1yr), while that of V1515 Cyg is
definitely longer ($t_{\mathrm{rise}}\approx$ 20 yrs). On the other
hand, while FU Ori and V1515 Cyg have a very long decay timescale
($t_{\mathrm{decay}}\approx 50-100$ yrs), V1057 Cyg decays much faster
($t_{\mathrm{decay}}\approx 10$ yrs).

Several different mechanisms have been suggested to trigger the
outburst: these include a tidal interaction with a companion star
\citep{bonnell92}, a gravitational instability in the outer, massive
disc \citep{armitage2001}, and a viscous-thermal instability
\citep[hereafter BL94]{bellin94} in a disc fed at a high enough mass
accretion rate from a surrounding envelope. In particular, the latter
model (which had been extensively studied in order to explain the
dwarf novae outbursts in binary systems, \citealt{meyer81}) has been
discussed in some details and is considered to be the most promising
model for the outburst.

Despite its ability to reproduce many features of the outburst, the
thermal instability model \citepalias{bellin94} has found some
difficulties in explaining the details of the FU Orionis light
curves. In fact, a thermal instability is triggered when the disc
surface density becomes larger than a critical value $\Sigma_A$,
dependent on the relevant opacities in the quiescent phase (see
Sect. \ref{sec:inst} below). In the \citetalias{bellin94} models this
occurs first very close to the inner edge of the disc, so that the
instability propagates inside-out, producing a slowly rising
luminosity. \citetalias{bellin94} indeed find rise timescales $\gtrsim
10$ yrs, which may be compatible with the light curve of V1515 Cyg,
but are too long to explain the rapid rise of FU Ori and of V1057 Cyg.
On the other hand, a rapid rise can be achieved if the instability is
first triggered at a large radius, so that the instability propagates
outside-in. \citet{clarkelin90} and \citet{belletal95} have
constructed models where the instability is triggered by introducing
some {\it ad hoc} density perturbations at $R\gtrsim 10 R_{\odot}$
from the central object, and succeeded in obtaining rapid rise
outbursts.

An alternative (and self-consistent) way to obtain rapid rise
outbursts was proposed by \citet{clarkesyer96}. They showed that the
presence of a companion with mass larger than the local disc mass (for
example, a massive planet) embedded in the disc can substantially
modify the disc structure, banking up the disc material just outside
the location of the companion. In this way, the banked up disc surface
density may eventually become larger than the critical value and the
instability can be triggered at the radius corresponding to the
position of the companion.

\citet{clarkearmi03} have recently suggested that a planet embedded in
the disc of an FU Orionis object would lead to a clear spectroscopic
signature in the form of a periodic modulation of the double-peaked
line profiles observed in these systems, with periods corresponding to
the orbital frequency of the planet. Interestingly, periodic
modulations in the line profiles have indeed been observed in a long
term monitoring campaign of the rapid rise FU Orionis objects V1057
Cyg and FU Ori \citep{herbig03}. The persistence of the feature (with
a period of $\approx$ 3 days) in successive observing series argues
against its origin as a transient disc surface feature, and is
consistent with the hypothesis of an embedded hot Jupiter.

In this paper we develop a simple, one-dimensional, time-dependent
outburst model following the suggestion of \citet{clarkesyer96} that
the outburst is triggered by the presence of a planet embedded in the
disc. We take into account the details of the planet-disc interaction
and of the thermal evolution of the disc. In this way we are able to
follow at the same time both the orbital evolution of the planet, the
banking up of disc material outside the planet position and the
details of the outburst phase. We indeed obtain rapid rise outbursts,
with the rise timescale and the location of the radius at which the
disc is first triggered into outburst depending mainly on the planet
mass. [Note that in this paper we are interested in the dynamical
effects of the presence of a companion with small mass ratio embedded
in the disc, to which we will refer as a ``planet'', even if we will
allow its mass to be above the nominal minimum mass for deuterium
burning, sometimes used to define planets as opposed to brown dwarfs.]

The paper is organised as follows: in Section \ref{sec:inst} we
introduce the thermal instability model and discuss the specific disc
model that we have adopted here. In Section \ref{sec:planet} we
discuss the details of the planet-disc interaction. In Section
\ref{sec:outburst} we show the results of our outburst simulations. In
Section \ref{sec:discussion} we discuss our results and in Section
\ref{sec:conclusions} we draw our conclusions.

\section{Thermal instability model}
\label{sec:inst}

\subsection{General features}
\label{instability}

The time evolution of an axisymmetric, thin, Keplerian accretion disc
(in which, for the moment, we do not include any effects coming from
the tidal interaction with an embedded planet) is determined by the
diffusion equation \citep{pringle81}:
\begin{equation}
\label{eq:diffusion}
\frac{\partial\Sigma}{\partial t}=\frac{3}{R}\frac{\partial}{\partial
  R} \left(R^{1/2}\frac{\partial}{\partial R}(R^{1/2}\mu)\right),
\end{equation}
where $R$ is the cylindrical radius, $\Sigma$ is the disc surface
density, and $\mu=\nu\Sigma=\mdot/3\pi$ is the mass flux at radius
$R$, where $\nu$ is the viscosity coefficient. The viscosity is
usually described in terms of some empirical prescription, such as the
$\alpha$ prescription \citep{shakura73}, according to which
$\nu=\alpha c_s H$, where $c_s$ is the disc thermal speed and $H$ is
the disc thickness. The value of $\mu$ can be obtained from the
solution of the energy balance equation for the disc, once a
prescription for the opacity (which determines the relevant radiative
cooling term) is given. In this paper, we will use the following
simplified form of the time-dependent energy balance equation
\citep{pringle86}:
\begin{equation}
\label{eq:thermal}
\frac{\partial\mu}{\partial t}=-\frac{\mu-\mu_{\mathrm{eq}}}{t_{\mathrm
    {therm}}} +\frac{\mu}{\Sigma}\frac{\partial\Sigma}{\partial t},
\end{equation}
where $t_{\mathrm{therm}}=\alpha^{-1}\Omega^{-1}$ is the thermal
time-scale, and $\Omega$ is the angular velocity of the disc. In
equation (\ref{eq:thermal}), $\mu_{\mathrm{eq}}$ is the value of $\mu$ at
thermal equilibrium, and is determined by the specific choice for the
opacity. In equation (\ref{eq:thermal}) we have neglected some terms,
namely the radial radiative flux (which is going to be of the order of
$H/R$ with respect to the other terms, and therefore negligible for a
thin disc) and the advective terms (i.e. terms in the form $\mathbf
u\cdot\nabla$, where $\mathbf u$ is the fluid velocity). The
simplified description adopted here is able to capture the main
properties of the outburst (see, for example, \citealt{pringle86}),
even if some quantitative result may actually be influenced by the
terms that we have neglected (see Section \ref{sec:discussion} below).

\begin{figure}
\centerline{\epsfig{figure=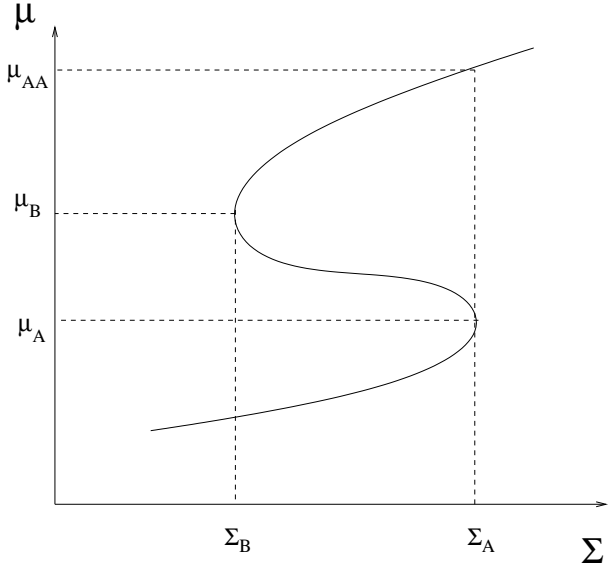,width=80mm}}
\caption{Typical S-curve in the $\Sigma$-$\mu$ plane at a given
  radius. $\Sigma_{\mathrm{A}}$ is the maximum disc density on the lower
  branch, $\Sigma_{\mathrm{B}}$ is the minimum disc density on the
  upper branch.}
\label{fig:scurve}
\end{figure} 

A given equilibrium solution is unstable if
$\partial\mu_{\mathrm{eq}}/\partial\Sigma<0$.  This happens, for
example, when the temperature of the disc is close to $10^4$K and
hydrogen becomes partially ionised \citepalias{bellin94}.  The
behaviour of a thermally unstable disc is best understood in the
$\Sigma$-$\mu$ plane at a fixed radius, where the equilibrium curve
can have a characteristic S-shape, in which two stable solutions are
connected by an intermediate unstable branch (see
Fig. \ref{fig:scurve}). The two critical values for the density,
$\Sigma_{\mathrm{A}}$ and $\Sigma_{\mathrm{B}}$, represent the maximum
disc density on the lower branch and the minimum disc density on the
upper branch, respectively. If, for some reason, the disc density
becomes greater than $\Sigma_{\mathrm{A}}$ the disc cannot stay in the
lower stable branch and jumps (on a thermal timescale) to the upper
branch, characterised by a much larger accretion rate. As a result of
the enhanced accretion rate, the surface density will decrease and
eventually become smaller than $\Sigma_{\mathrm{B}}$, and the disc
will go back to the lower quiescent branch. Of course, the process
described above refers to an isolated disc annulus, but when the
instability is triggered somewhere in the disc it propagates to the
adjacent annuli with a typical velocity $\approx \alpha
c_{\mathrm{s}}$ \citep{LPF85}. As a result, when the instability
propagates inside-out, since the thermal velocity of the disc
decreases with increasing radius, it will slow down during the
propagation, producing a slowly rising light curve. In addition,
viscous diffusion always results in a higher mass fraction evolving
inward and this, together with the fact that in general
$\Sigma_{\mathrm{A}}$ decreases at small radii (see below) makes the
inward front propagation faster.

The time spent by the disc on the upper branch can be estimated simply
as the viscous time-scale at the outer radius where the instability
propagates: 
\begin{equation}
\label{eq:duration}
t_{\nu}\approx\frac{R_{\mathrm{lim}}^2}{\nu(R_{\mathrm{lim}})}.
\end{equation}

\subsection{Reference disc model and numerical method}

As we have seen in the previous Section, to set up a thermal
instability model we need to specify the details of the equilibrium
curves (the S-shaped curve in Fig. \ref{fig:scurve}) at each
radius. In this work we will use approximate parameterisations of the
detailed vertical structure calculations by \citetalias{bellin94}:
\begin{eqnarray}
\label{eq:low}
\nonumber
\Sigma_{\mathrm{low}}=5.3~10^6\left(\frac{\mdot}{10^{-6}\msunyr}\right)^{0.8} 
\left(\frac{R}{10R_{\sun}}\right)^{-1}\\
\left(\frac{\alpha}{10^{-4}}\right)^{-0.8}\mbox{g/cm$^2$},
\end{eqnarray}
\begin{eqnarray}
\label{eq:high}
\nonumber
\Sigma_{\mathrm{high}}=1.1~10^5\left(\frac{\mdot}{10^{-6}\msunyr}\right)^{0.75}
\left(\frac{R}{10R_{\sun}}\right)^{-0.85}\\
\left(\frac{\alpha}{10^{-4}}\right)^{-0.8}\mbox{g/cm$^2$},
\end{eqnarray}
\begin{equation}
\label{eq:sigmaA}
\Sigma_{\mathrm{A}}=1.6~10^6\left(\frac{R}{10R_{\odot}}\right)^{1.1}
\left(\frac{\alpha}{10^{-4}}\right)^{-0.8} \mbox{g/cm$^2$},
\end{equation}
\begin{equation}
\label{eq:sigmaB}
\Sigma_{\mathrm{B}}=6.7~10^5\left(\frac{R}{10R_{\odot}}\right)^{1.1}
\left(\frac{\alpha}{10^{-4}}\right)^{-0.8} \mbox{g/cm$^2$},
\end{equation}
where $\Sigma_{\mathrm{low}}$ and $\Sigma_{\mathrm{high}}$ are the
cold and hot stable branches of the equilibrium curve, respectively,
and where $\Sigma_{\mathrm{A}}$ and $\Sigma_{\mathrm{B}}$ are the
critical values of the density (see Fig. \ref{fig:scurve}). We will
also assume that the unstable branch is a straight line in the plane
$\log\Sigma$-$\log\mu$, connecting the points $(\Sigma_{\mathrm{A}}$, 
$\mu_{\mathrm{A}})$ and $(\Sigma_{\mathrm{B}}$, $\mu_{\mathrm{B}})$
(see Fig. \ref{fig:scurve}).

In thermal instability models of dwarf novae outbursts (see
\citealt{lasota01} for a review) it has become customary to adopt a
larger value of the viscosity parameter $\alpha$ in the upper branch
with respect to the lower branch, in order to correctly reproduce the
light curves of the outburst. This might have some physical
justification in the scenario in which the main mechanism for angular
momentum transport in the disc is MHD turbulence: in fact, in the
lower branch the disc has a lower degree of ionisation with respect to
the higher branch, so that magnetic instabilities are expected to be
less effective in promoting accretion. In this work, we will conform to
this practice and will adopt two different values for $\alpha$ in the
two equilibrium branches. Our reference values are
$\alpha_{\mathrm{low}}=10^{-4}$, $\alpha_{\mathrm{high}}=10^{-3}$, but
we have also explored different choices for $\alpha$ (see below). The
transition from the low to the high value of $\alpha$ is followed
smoothly when the disc moves from the lower to the upper branch and
viceversa.

We have solved the coupled equations (\ref{eq:diffusion}) and
(\ref{eq:thermal}) using an explicit scheme, with a radial grid
equally spaced in the variable $R^{1/2}$. The inner and outer disc
radii are taken to be $R_{\mathrm{in}}=3R_{\odot}$ and
$R_{\mathrm{out}}=150R_{\odot}$, respectively. We have taken a free
flow inner boundary condition and a reflecting outer boundary
condition. 

In all our outburst models we have fed the disc at an accretion rate
$\mdot_{\mathrm{in}}$ kept constant with time. This is simply
implemented by adding a source term in equation (\ref{eq:diffusion}). This
term is taken to be vanishing outside a narrow radial range close to
the outer boundary of the disc.

To save computational time, if all the disc lies on the lower branch
(and if $\Sigma$ is everywhere smaller than $\Sigma_{\mathrm{A}}$),
rather than solving equation (\ref{eq:thermal}), we simply assume that
$\mu$ is equal to its equilibrium value on the lower branch, which is
equivalent to the requirement that the density at each point satisfies
equation (\ref{eq:low}). In this way the limiting factor for the time-steps
in our code is simply the stability condition for a viscous flow,
which generally gives a time-step much longer than the thermal
time-scale on which we should evolve the code if we also solved
equation (\ref{eq:thermal}).

\subsection{Simulated inside-out outbursts}

As a test of our numerical scheme, we have first simulated some
inside-out outbursts of the same kind of those obtained by
\citetalias{bellin94}, without any perturbation (neither artificially
imposed, as in \citealt{clarkelin90}, nor self-consistently obtained
through the presence of a planet).

\begin{figure*}
\centerline{ \psfig{figure=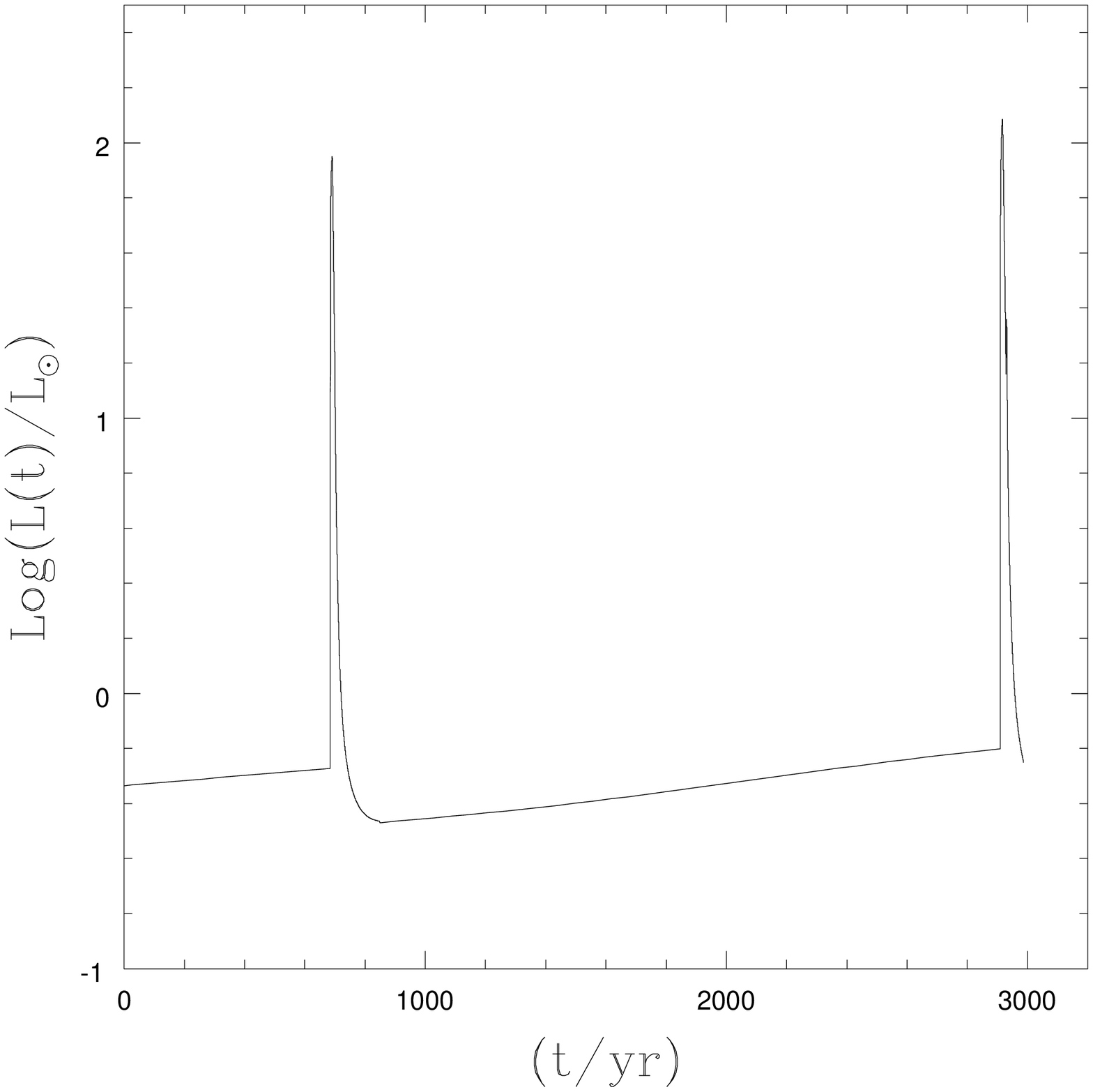,width=.5\textwidth}
             \psfig{figure=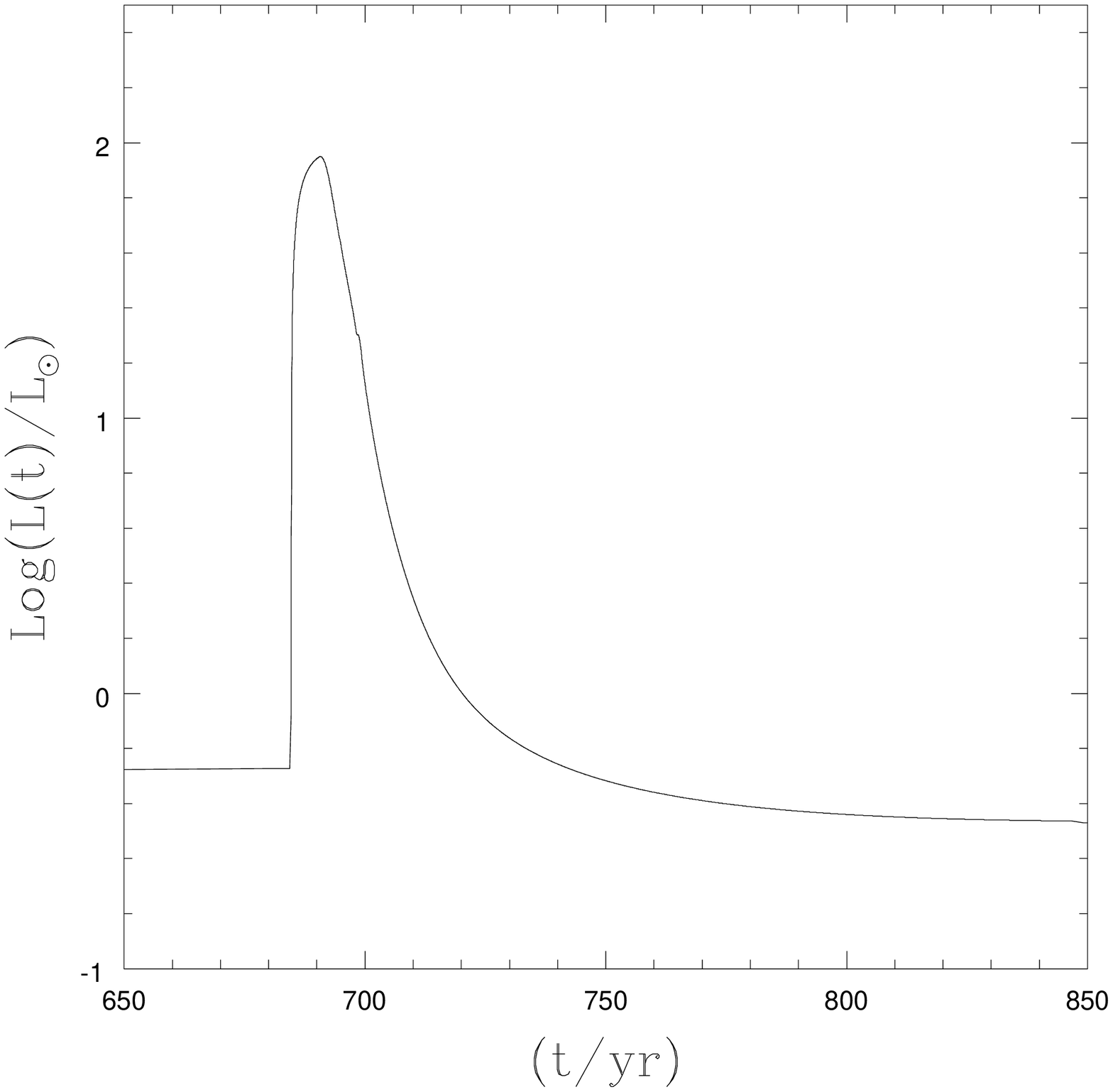,width=.5\textwidth}}
\caption{Light curves of simple inside-out outbursts. In this case,
  the imposed mass accretion rate was
  $\mdot_{\mathrm{in}}=3~10^{-6}\msunyr$. The right panel shows a blow
  up of the outburst light curve.}
\label{fig:noplan}
\end{figure*}

Fig. \ref{fig:noplan} shows the light curves obtained in this way, for
an imposed accretion rate $\mdot_{in}=3~10^{-6}\msunyr$ (so as to be
consistent with \citetalias{bellin94}). The central object mass is
assumed to be $M_{\star}=M_{\odot}$ (we will use this assumption
throughout the paper).  We are indeed able to obtain repetitive
outbursts, with a duration of $\approx 100$ yr and a recurrence time
of $\approx 2000$ yr, in fair agreement with
\citetalias{bellin94}. The peak of the bolometric luminosity is
$L_{\mathrm{peak}}\approx 10^2L_{\odot}$, and the mass accretion rate
during the outburst is $\mdot\approx 6~10^{-5}\msunyr$. The rise in
luminosity is characterised by an initially rapid increase, which
eventually slows down so that the maximum is achieved after $\approx
10$ yr, in agreement with the expectations from an inside-out
outburst. The outburst is indeed an inside-out outburst, the radius
where the instability is first triggered being $\approx
4R_{\odot}$. The outer radius to which the instability propagates in
this case was $R_{\mathrm{lim}} \approx 25R_{\odot}$, again in
agreement with the results of \citetalias{bellin94}.

In other simulations, not shown here, we have also been able to
reproduce the variation of the outburst features with respect to
parameters variations (different choices of $\alpha$, different values
of $\mdot_{\mathrm{in}}$) as described in \citetalias{bellin94}. The
agreement that we obtain is remarkable, considering that our outburst
model is significantly simplified with respect to the one used by
\citetalias{bellin94}. In particular, ({\it i}) while we use simple
parameterisations for the equilibrium curves, \citetalias{bellin94}
solve explicitly in much greater detail the vertical disc structure
and the radiative transfer, including also non-thermal-equilibrium
computations; ({\it ii}) \citetalias{bellin94} retain all the
advective terms and the radial energy fluxes in their thermal
equation, while we neglect them. Our results show that the main
qualitative features of the outburst are not significantly affected by
these more refined calculations.

\section{Planet-disc interaction}
\label{sec:planet}

In this Section we will leave for a while the subject of thermal
instability in discs and will concentrate on the second basic
ingredient of our model, i.e. the interaction between an accretion
disc and an embedded satellite. It is worth noting that the
calculations described in this Section can be applied also to similar
systems on a much larger scale, such as the case of a binary of two
supermassive black holes \citep{natarajan02}.

\subsection{Generalities: gap formation, planet migration}

The topic of disc-satellite interaction and the connected issue of
satellite migration has been explored in a number of papers
\citep[e.g.,][]{linpap79b,linpap79a,goldreich80}. It is well known
that a massive enough satellite is going to open up an annular gap in
the disc at its orbital radius as a result of the gravitational
torques exerted between the satellite and the disc. This angular
momentum exchange will also cause the satellite to migrate to smaller
radii, in the so-called Type II migration (as opposed to the much
faster Type I migration which occurs when the planet mass in not high
enough to open up a gap in the disc).

In this paper we will use a simplified expression for the tidal
torque, obtained in the ``impulse approximation'', by analogy with the
process of dynamical friction. In this case, equation (\ref{eq:diffusion})
should be modified as follows:
\begin{equation}
\label{eq:diffplanet}
  \frac{\partial\Sigma}{\partial
  t}=\frac{3}{R}\frac{\partial}{\partial R}
  \left(R^{1/2}\frac{\partial}{\partial R}(R^{1/2}\mu)\right)
  -\frac{1}{R}\frac{\partial}{\partial
  R}\left(\frac{\Lambda\Sigma}{\Omega}\right),
\end{equation}
where the specific tidal torque $\Lambda$ is given by:
\begin{eqnarray}
\label{eq:torque}
\Lambda= & q^2\Omega^2R^2\left(\displaystyle
\frac{R_{\mathrm{s}}}{p}\right)^4  &
R>R_{\mathrm{s}} \\
\nonumber \Lambda= & -q^2\Omega^2R^2\left(\displaystyle 
\frac{R}{p}\right)^4  & R<R_{\mathrm{s}}. 
\end{eqnarray}
In equation (\ref{eq:torque}), $R_{\mathrm{s}}$ is the radial position
of the satellite, $q=M_{\mathrm{s}}/M_{\star}$ is the mass ratio
between the satellite and the central object, and
$p=R-R_{\mathrm{s}}$. This simplified form of the specific torque is
the same as commonly used by many investigators (see, e.g.,
\citealt{armitage2001,armibonnel2002}).  This formalism, although
simplified, has been shown to provide results in agreement with more
sophisticated numerical treatments \citep{trilling98}. In addition,
for the massive planets that we are considering here, the form of the
torque plays a minor role in the resulting dynamics (at least during
the quiescent phase of the limit cycle), since the satellite is
massive enough to produce a clear gap. In such Type II migration, the
evolution of the planet is insensitive of the form of the torque
because the disc surface density profile self-adjusts so as to lock
the planet orbital migration to the evolution of the edge of the
gap. The specific form of the torque will be more important during the
outburst phase, when the gap is going to be filled. In this case,
estimates of the torques (and of the timescale for Type I migration)
based on the linear theory \citep{tanaka02} might not be applicable
for a number of reasons: \uno the interaction with the high mass
planets that we consider here might be non-linear; \due these estimates
usually neglect the torques coming from inside the Roche radius of the
planets, which might be important \citep{bate03}; \tre runaway
migration (as described by \citealt{masset03}) might influence
significantly the results, when the disc mass is high. Most 2D and 3D
simulations of planet-disc interaction in this embedded case
\citep{lubow99,bate03,masset03} are not useful to our purposes, since
they generally explore the case where the disc is a much colder, T
Tauri-like disc, in which case a planet that does not open up a gap
needs to have a small mass (less than $1M_{\mathrm{Jupiter}}$).
Detailed numerical simulations of high mass planets embedded in hot
discs would be needed to clearly assess this problem (incidentally,
such simulations would also have the chance of better resolving the
region within the Roche lobe of the planet, since this is going to be
relatively large). Such analysis is, however, beyond the immediate
goals of the present paper. In the absence of clear results, we have
therefore preferred to use the above formulation for the torque also
during the outburst.

For numerical reasons, we have smoothed the torque term when $R\approx
R_{\mathrm{s}}$, where the torque would have a singularity (see
equation (\ref{eq:torque})). We have used the same smoothing
prescription as in \citet{clarkesyer95} and \citet{linpap86b}, so that
the torque is smoothed when $|R-R_{\mathrm{s}}|
<\max[H,R_{\mathrm{H}}]$, where $H$ is the disc thickness and
$R_{\mathrm{H}}$ is the size of the Hill sphere of the planet (note
that, for the disc models and planetary masses considered here, during
quiescence generally $R_{\mathrm{H}}>H$, while during the outburst $H$
and $R_{\mathrm{H}}$ are comparable and of the order of $\approx
0.1R$).

The back reaction of the disc on the planet orbital motion can be
obtained by simply imposing angular momentum conservation, that can be
written in the form:
\begin{equation}
\label{eq:planet}
\frac{\de}{\de t}(M_{\mathrm{s}}\Omega_{\mathrm{s}}R_{\mathrm{s}}^2)=
-\int_{R_{\mathrm{in}}}^{R_{\mathrm{out}}}2\pi R\Lambda\Sigma\de R,
\end{equation}
where the integral is taken over the whole disc surface.

The behaviour of the coupled disc-satellite system depends on two
dimensionless parameters:
\begin{equation}
A=\frac{\Omega R^2}{\nu}q^2,
\end{equation}
measures the relative strength of the second and first terms on the
right-hand-side of equation (\ref{eq:diffplanet}), while:
\begin{equation}
B=\frac{4\pi\Sigma R^2_{\mathrm{s}}}{M_{\mathrm{s}}}
\end{equation}
gives a measure of the magnitude of the right-hand-side of
equation (\ref{eq:planet}). 

The parameter $A$ has a simple physical interpretation. In fact, it
can be shown \citep{linpap79a} that $A=(\Delta/R)^3$, where $\Delta$
is the gap width. In order to open up a gap, the gravitational effect
of the satellite has to overcome the pressure of the disc and the
viscosity, that both act as to counteract the gap
opening. \citet{takeuchi96} have shown that a gap is going to be open
when $q\gtrsim 2\alpha^{1/2}(H/R)^2$.

The second important parameter, $B$, represents the relative local
disc mass with respect to the satellite. If the satellite is more
massive than the disc (i.e. $B\ll 1$) its inertia will cause a rather
slow migration, whereas in the opposite case ($B\gg 1$) the planet
will migrate on the local viscous timescale, behaving like a
representative fluid element of the disc. In the limit where $B\ll 1$
the satellite will act like a dam for the flow, preventing accretion
through the gap, and will cause the upstream surface density to bank
up \citep{clarkesyer95}. In general, $B$ is an increasing function of
radius. We can therefore define a radius $R_{\mathrm{B}}$, at which
$B=1$, so that for $R\gg R_{\mathrm{B}}$ the planet will be light
compared to the disc and it will simply migrate with the accretion
flow. For example, for the quiescent surface density profile that we
adopt here (equation (\ref{eq:low})), we have:

\begin{eqnarray}
\nonumber
R_{\mathrm{B}}\approx 0.62 R_{\odot}\left(\frac{\alpha}{10^{-4}}\right)
^{0.8}\left(\frac{M_{\mathrm{s}}}{M_{\mathrm{Jupiter}}}\right)\\
\left(\frac{\mdot}{10^{-6}\msunyr}\right)^{-0.8}.
\label{eq:rb}
\end{eqnarray}

\begin{figure*}
\centerline{ \psfig{figure=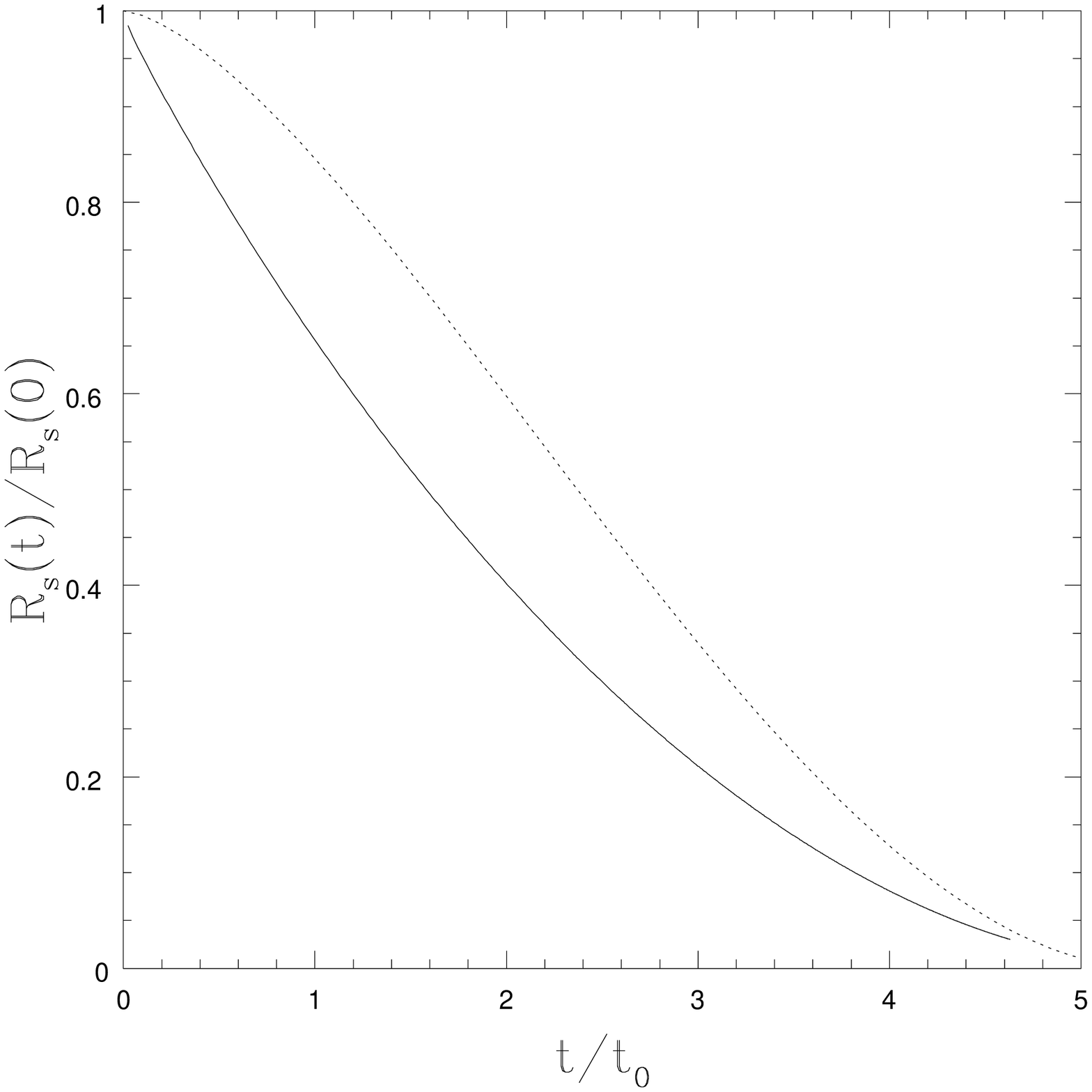,width=.5\textwidth}
             \psfig{figure=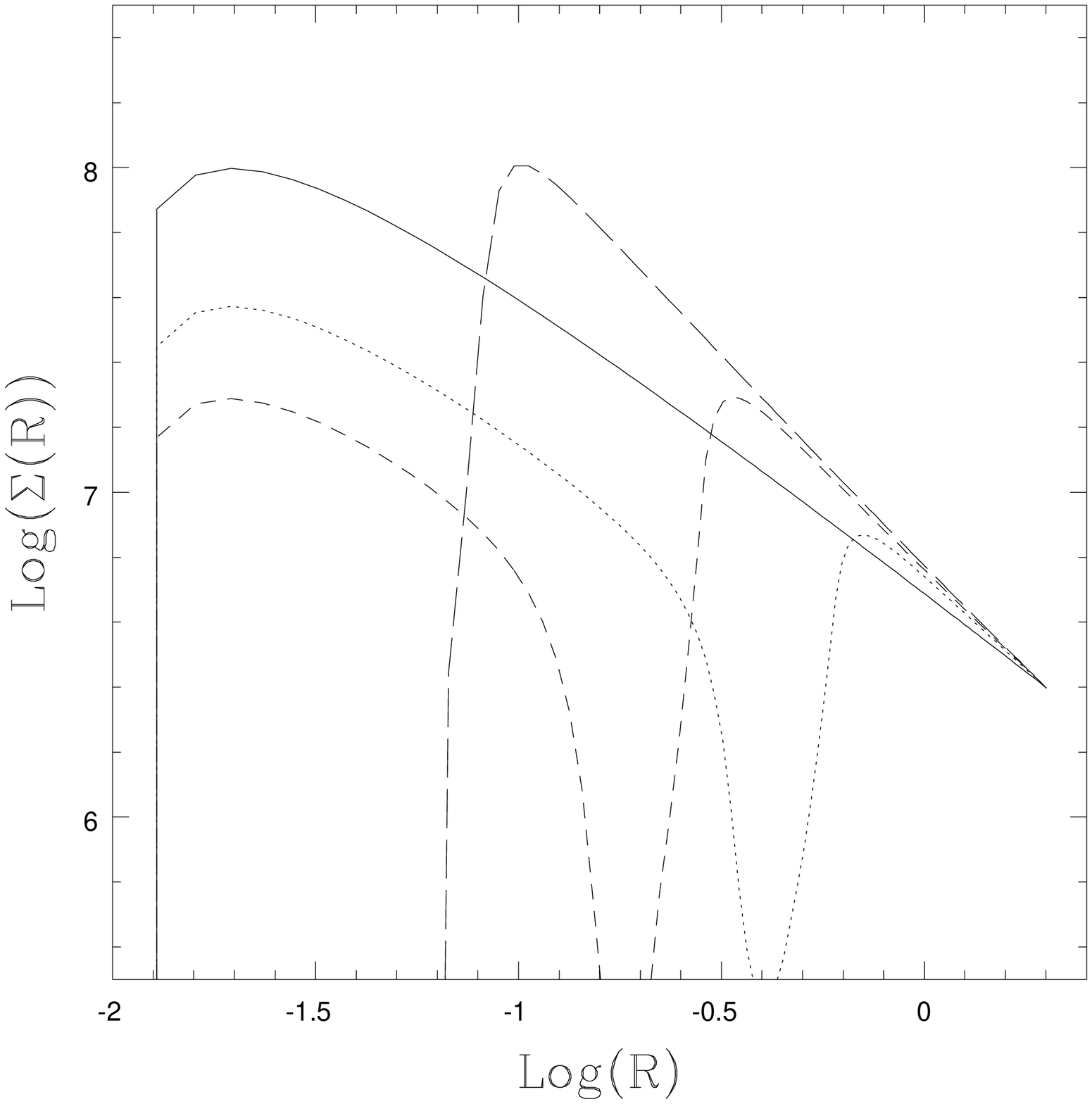,width=.5\textwidth}}
\caption{Left panel: Evolution of the satellite orbit in our numerical
  simulation (solid line) and according to the analytical solution of
  \citet{ivanov99} (dotted line). Right panel: Initial surface density
  profile (solid line), compared with the surface density profile of
  the disc at three different times, i.e. $t/t_0=1.7,~2.9,~4.1$,
  respectively.}
\label{fig:ivanov}
\end{figure*}

\subsection{Simulated planet migration}

\citet{clarkesyer95} and \citet{ivanov99} have found analytical
solutions for the orbital evolution of the planet and for the banking
up of surface density in the disc in the approximation in which the
gap width is assumed to be very small and equation (\ref{eq:diffplanet}) is
substituted by an appropriate boundary condition at the planet
position. In these computations the inner disc (i.e. the disc inside
the satellite orbit) is assumed to be empty, and initially the outer
disc is assumed to have the surface density corresponding to an
equilibrium disc accreting at a constant rate. 

As we have already mentioned, the evolution of the system (concerning
the satellite-disc evolution) is ``scale-free'' and is determined only
by the two parameters $A$ and $B$, and by the functional relation
between $\Sigma$ and $\mu$, which is normally taken to be in the form
$\Sigma\propto \mu^a R^b$ (cf. equations (\ref{eq:low}) and
(\ref{eq:high})). If initially the disc is in a steady state at some
given accretion rate $\mdot$, the satellite evolution, according to
\citet{ivanov99}, is given by:
\begin{equation}
\label{eq:ivanov}
R_{\mathrm{s}}=R_{\mathrm{s0}}[1-\gamma B_0 (t/t_0)^d]^2,
\end{equation}
where $R_{\mathrm{s0}}$ is the initial position of the planet, and
$\gamma$ and $d$ are two dimensionless parameter of order of unity,
dependent only on $a$ and $b$. $B_0$ is the initial value of $B$ (in
the unperturbed disc) and $t_0$ is the viscous timescale in the
unperturbed disc at $R=R_{\mathrm{s0}}$:
\begin{equation}
t_0=\frac{4R_{\mathrm{s0}}^2}{3\nu(R_{\mathrm{s0}})}.
\end{equation}

In this work we have decided to take a different approach with respect
to \citet{ivanov99} and to solve numerically equations
(\ref{eq:diffplanet}) and (\ref{eq:planet}). In fact, ({\it i}) in the
outburst simulations that we will present in Section
\ref{sec:outburst} we will not have an initially steady state disc,
but we will feed it from outside at a given rate, and ({\it ii}) one
important goal of this work will also be to check what surface density
profile will be left over in the inner disc after the outburst, and
whether the satellite will still be able to open a gap at that
stage. Therefore, we will need to treat in full details the density
evolution of the disc at all radii.

We have used the same explicit code discussed above in Section
\ref{sec:inst}, and the gravitational torque has been computed using
the Lelevier method \citep{potter}. As mentioned above, we have
smoothed the torque term in equation (\ref{eq:torque}) . This
smoothing will not affect the overall evolution of the satellite, but
is going to prevent the gap from being fully open, so that a small
leakage (with a magnitude of a few percent of the unperturbed
accretion rate) from the outer to the inner disc is present in our
simulations. Three-dimensional simulations of planet-disc interactions
\citep{lubow99,bate03} often show that a leakage through the gap is
actually occurring, so that our model is not going to be
unrealistic. This mass flow through the gap can be significant for low
mass planets, but in the case we are considering here, with a high
mass planet embedded in a cold, low-viscosity disc, a clear gap is
going to be open, and the mass flow through the gap is very small.
\citet{bryden2000} estimate the amount of mass flow through the gap as
a function of the relevant disc parameters, finding that:
\begin{eqnarray}
\nonumber 
\left.\frac{\de M}{\de t}\right|_{\mathrm{gap}}\approx 
5~10^{-6} q^{-1.5} \left(\frac{H}{R}\right)^{7.5}
\left(\frac{\Sigma}{200 \mathrm{g/sec}}\right)\\
\left(\frac{R_{\mathrm{s}}} {5.2 \mathrm{au}}\right)^{1/2} \msunyr.
\label{eq:flow}
\end{eqnarray}
During quiescence, the typical values for the disc density and
thickness can be evaluated from equation (\ref{eq:low}). We find that at a
typical radius of $R\approx 10R_{\odot}$, $H/R\approx
0.035$. According to equation (\ref{eq:flow}), the mass flow through the
gap is then going to be only $\approx 2.5~10^{-10}\msunyr$, consistent
with the numerical ``leakage'' that we allow in our simulations.

We have checked our planet-disc interaction numerical scheme
simulating the evolution of a satellite in an initially steady-state
disc (as in \citealt{ivanov99}), with no external feeding. The results
are shown in Fig. \ref{fig:ivanov}. In this simulation the initial
values of $A$ and $B$ were $A_0=0.1$ and $B_0=0.1$, respectively, and
the equilibrium curve is given by equation (\ref{eq:low}). With this
choice for the surface density profile, $B$ grows linearly with radius
(see equation (\ref{eq:rb})), so that
$R_{\mathrm{B}}=10R_{\mathrm{s0}}$. The left panel of
Fig. \ref{fig:ivanov} shows the orbital evolution of the satellite
(solid line), compared to the analytical prediction according to
equation (\ref{eq:ivanov}) (dotted line). The simulated migration
agrees reasonably well with the approximate analytical expectation,
being only slightly faster. (Note, however, that \citealt{ivanov99}
also numerically simulate the evolution of a satellite, within the
infinitesimal boundary layer model, and find as well a slightly faster
evolution with respect to the analytical formula).

The right panel in Fig. \ref{fig:ivanov} shows the initial surface
density of the disc (solid line), compared with the surface density at
three different times during the planet evolution. We can clearly see
the formation of the gap and the inward migration of the satellite
(cf. analogous results in \citealt{natarajan02} in the context of
supermassive black hole binaries). Most importantly for our study,
this plot shows clearly how the presence of the satellite is going to
enhance the upstream surface density of the disc with respect to its
initial value, and how the inner disc is rapidly depleted. In the next
Section we will show how, as a consequence of this ``banking up'' of
material, the upstream surface density can become larger than the
critical value for thermal instability $\Sigma_{\mathrm{A}}$, thus
triggering the outburst at a relatively large radius.

As a final comment, we note that, under particular circumstances, the
migration of the planet can be significantly modified by the effect of
corotation resonances, that we have not included here, and that might
lead to a rapid ``runaway'' migration \citep{masset03}. If the surface
density profile of the disc is extremely shallow (with $\Sigma$
falling off with radius less rapidly than $R^{-1/2}$) this type of
migration might even be outward. However, for these effects to be
important, the gap should not be completely empty, so that the
corotation region would contain enough mass to provide a significant
torque. Indeed, \citet{masset03} only find runaway migration to occur
in the transition region between Type I and Type II migration, when
the gap is partially open. For planet masses larger than
$1M_{\mathrm{Jupiter}}$ they recover the standard Type II migration
(see Fig. 12-14 in \citealt{masset03}). We therefore conclude that
during the quiescent phase corotation torques are not going to
significantly alter our picture.

\begin{figure*}
\centerline{ \psfig{figure=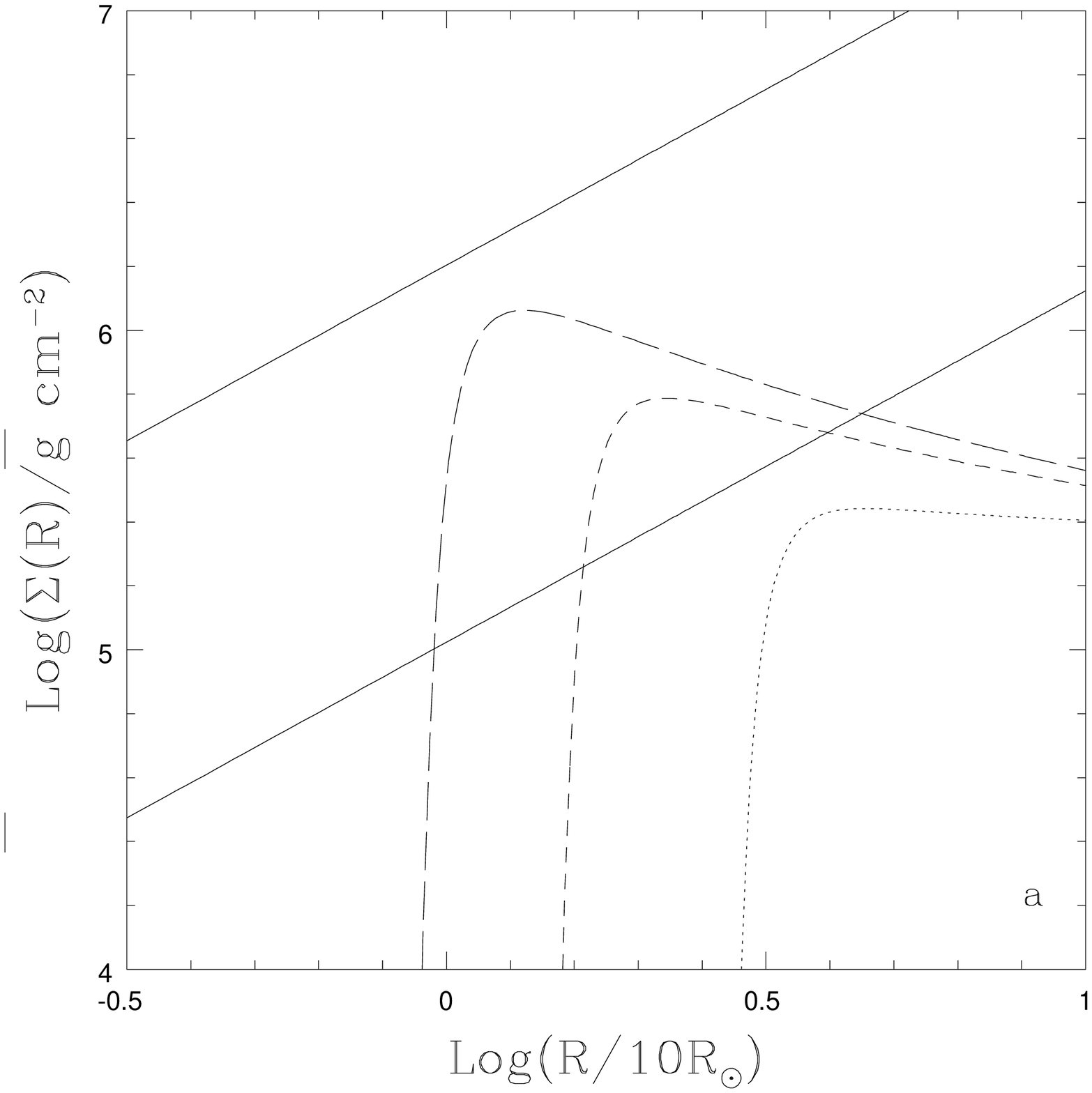,width=.4\textwidth}
             \psfig{figure=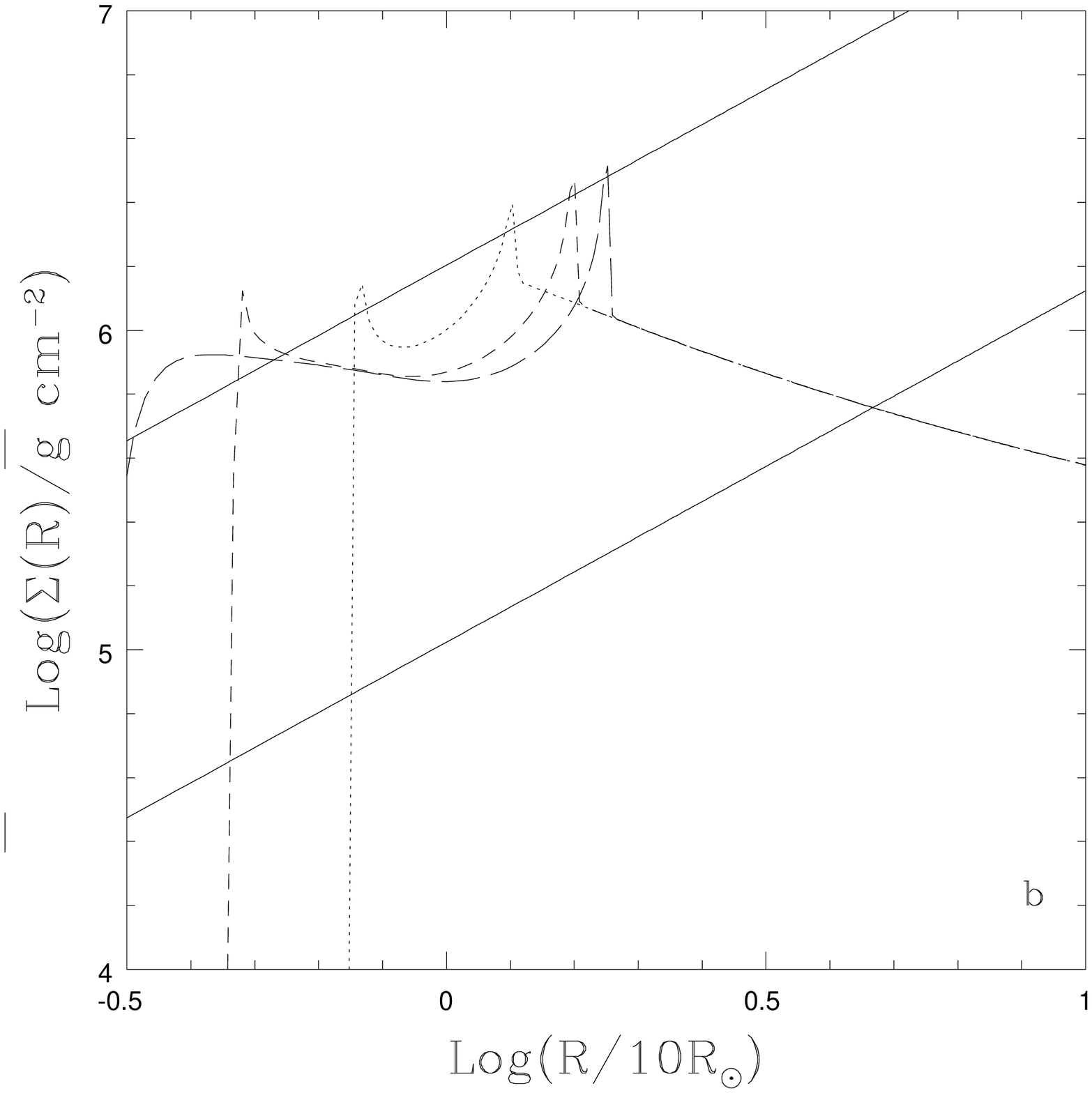,width=.4\textwidth}}
\centerline{ \psfig{figure=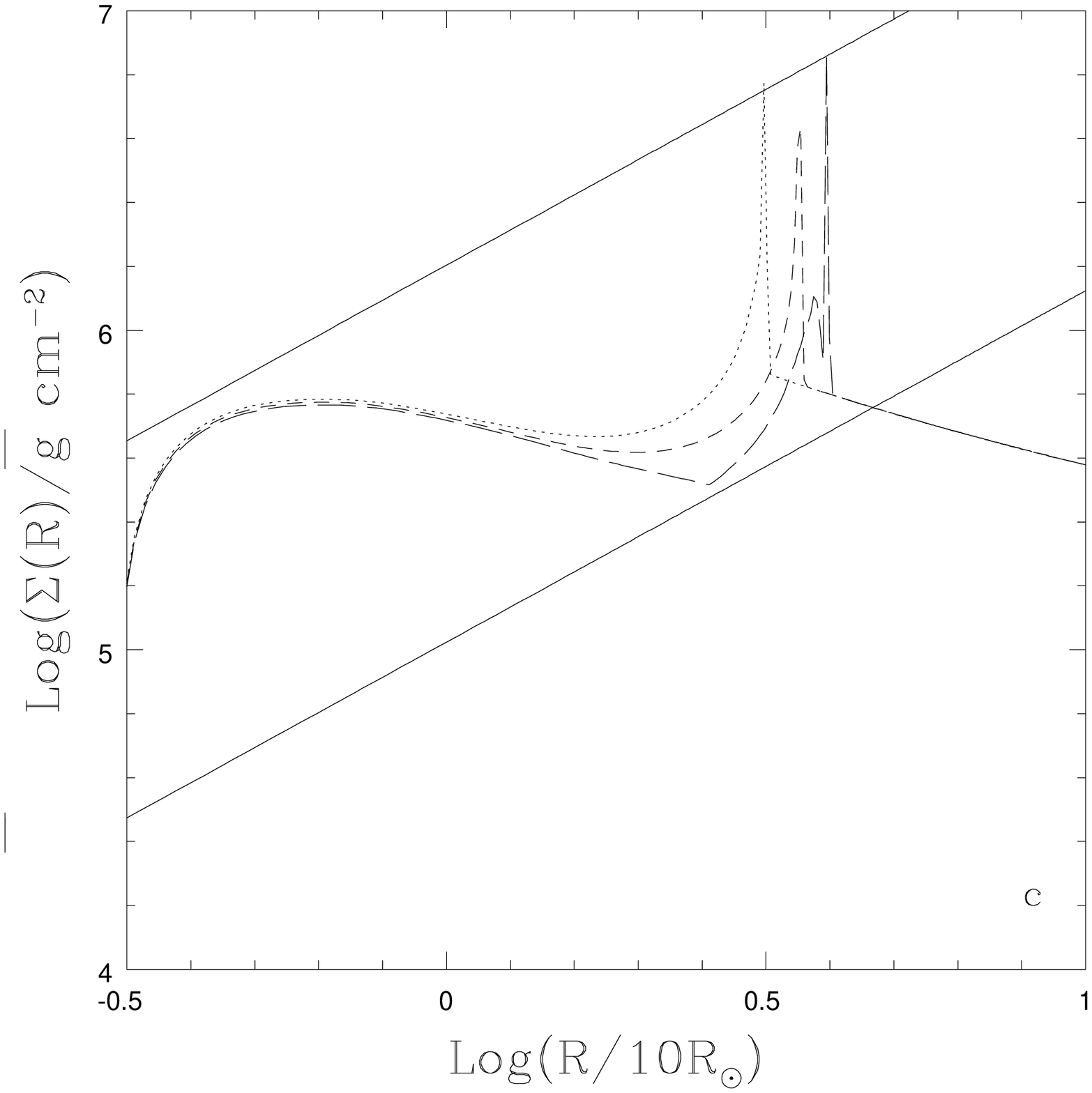,width=.4\textwidth}
             \psfig{figure=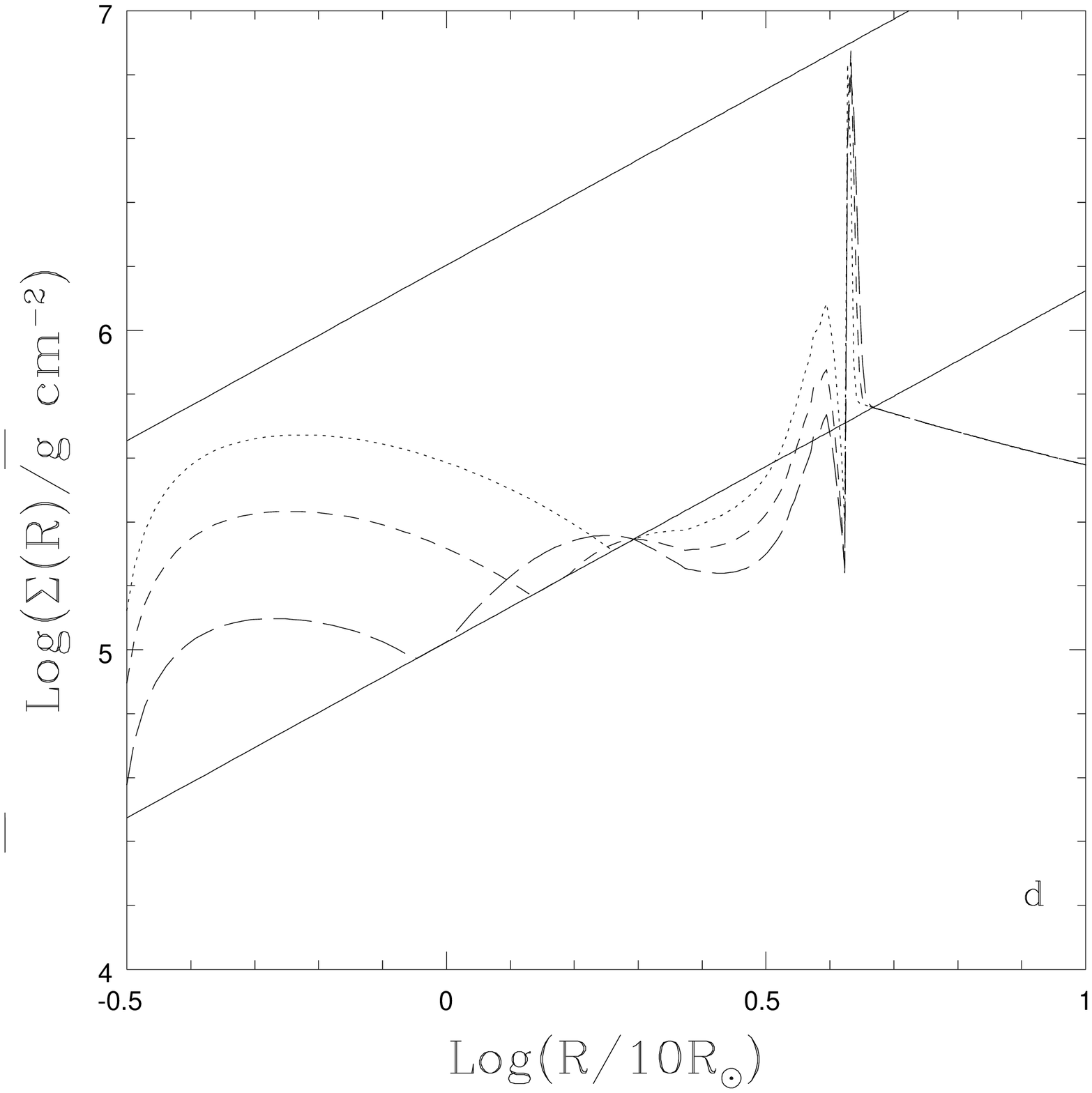,width=.4\textwidth}}
\centerline{ \psfig{figure=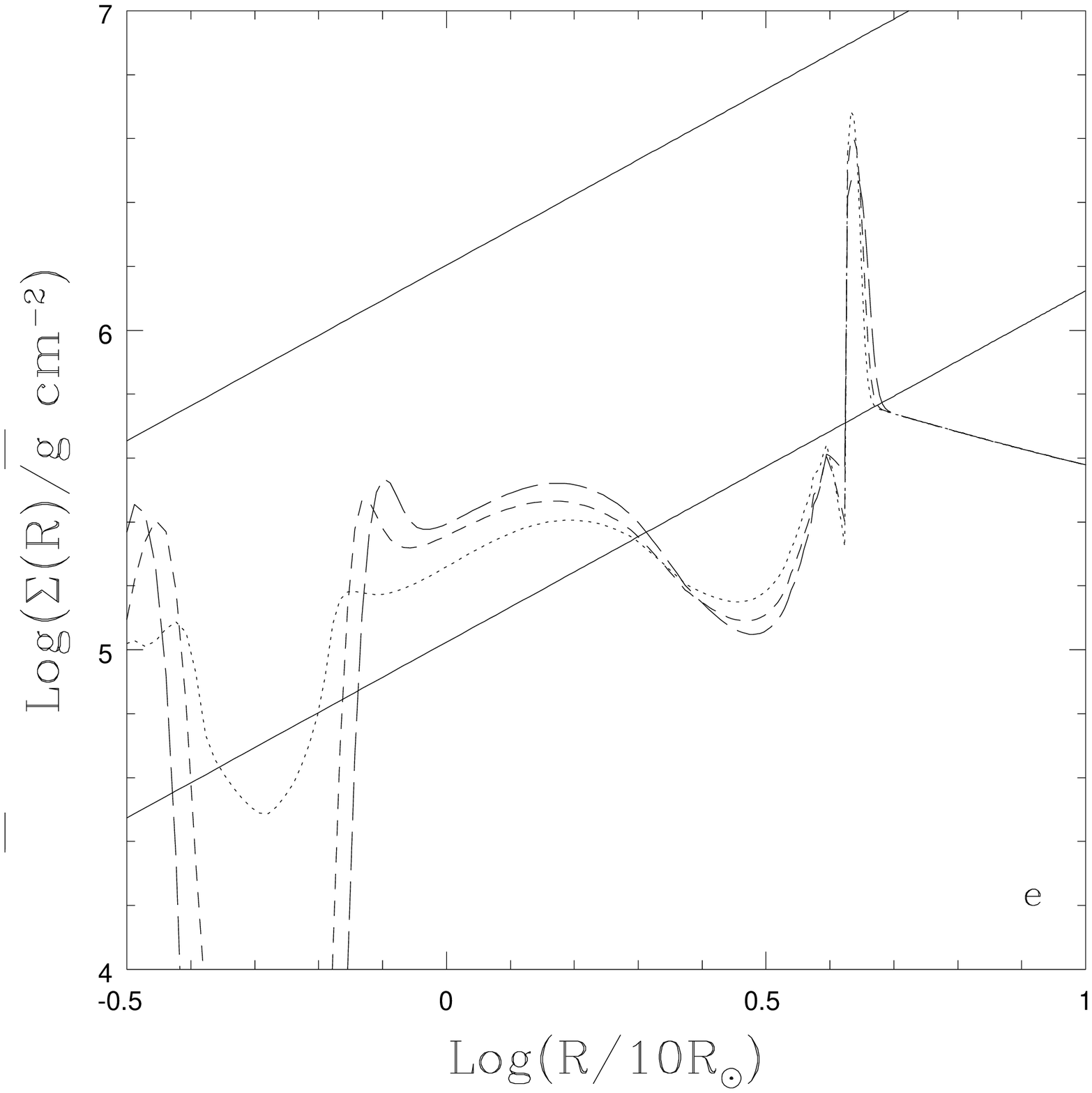,width=.4\textwidth}
             \psfig{figure=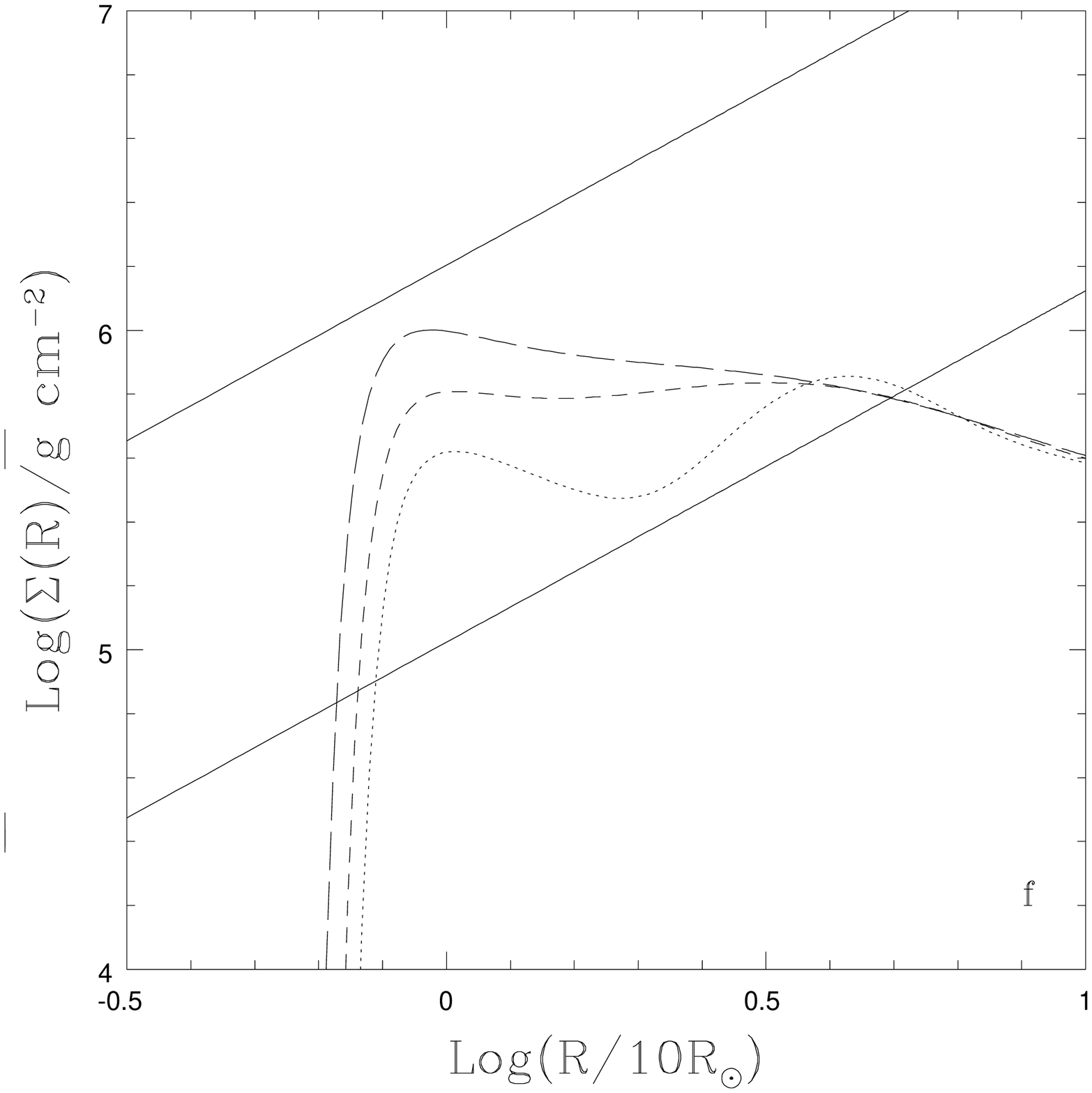,width=.4\textwidth}}

\caption{Evolution of the disc surface density profile during the
  outburst in the reference case. The two solid lines show the
  critical surface densities $\Sigma_{\mathrm{A}}(R)$ and
  $\Sigma_{\mathrm{B}}(R)$. (panel {\it a}: $t=20000-33000$ yrs) The
  surface density is banked up upstream of the planet position; (panel
  {\it b}: $t=35176-35178$ yrs) the outburst is triggered at $R\approx
  10R_{\odot}$ and the inward propagation front of the instability
  rapidly makes its way to the inner edge of the disc; (panel {\it c}:
  $t=35180-35190$ yrs) the outer propagation front slowly reaches the
  outer radius at which the instability can propagate; (panel {\it d}:
  $t=35200-35212$ yrs) the instability retreats; (panel {\it e}:
  $t=35250-35260$ yrs) the planet is able to re-open a gap; (panel
  {\it f}: $t=36000-38000$ yrs) the disc is ready for another
  outburst.}
\label{fig:reference}
\end{figure*}

\section{Planet induced outside-in outbursts}
\label{sec:outburst}

We have obtained planet-generated outbursts in FU Orionis objects by
combining the planet-disc interaction code described in Section
\ref{sec:planet} with the thermal evolution code described in Section
\ref{sec:inst}. In what follows we have always used the equilibrium
curves defined by equations (\ref{eq:low})-(\ref{eq:sigmaB}). All our
outburst simulations started with a relatively ``empty'' disc, i.e. a
steady state accretion disc on the lower equilibrium branch accreting
at relatively low rate, $\mdot_0=10^{-10}\msunyr$. This initially very
light disc is fed at a constant rate during the simulation, as
described in Section \ref{sec:inst}, but, contrary to that case, now
the matter which is added in the outer disc cannot flow freely to the
smallest radii because of the banking up effect due to planet.

In the next Subsection we will describe the details of the outburst
generated in our ``reference case'', in which the planet mass is
$M_{\mathrm{s}}=10M_{\mathrm{Jupiter}}$, the disc is fed at
$\mdot_{\mathrm{in}}=3~10^{-6}\msunyr$, and where
$\alpha_{\mathrm{low}}=10^{-4}$ and $\alpha_{\mathrm{high}}
=10^{-3}$. The initial orbital radius of the planet is
$40R_{\odot}$. The values of the input mass accretion rate and of
$\alpha$ are chosen so as to easily compare with the results of
\citetalias{bellin94}. In Subsection \ref{sec:parameter} we will
discuss the effects of changing the relevant parameters.

\subsection{Details of the reference case}

Fig. \ref{fig:reference} shows the various phases of the disc
evolution and of the outburst. In each plot the surface density of the
disc is plotted at three different times, along with the critical
values $\Sigma_{\mathrm{A}}(R)$ and $\Sigma_{\mathrm{B}}(R)$,
displayed with solid lines. Initially, the planet opens up a gap in
the disc (as one would expect from 3D simulations in the case of a
massive planet, cf. \citealt{bryden99,bate03}) and the inner disc is
rapidly drained into the central star, leaving an almost empty inner
``hole'', while the planet migrates inward (panel a). This first phase
occurs on the slow viscous timescale. The three snapshots in panel (a)
correspond to $t\approx$ 23000, 28500, and 33000 yrs,
respectively. When the surface density of the disc reaches
$\Sigma_{\mathrm{A}}$, the instability is triggered: in the reference
case this happens at $R\approx 10 R_{\odot}$; the region of the disc
where the instability is triggered moves to the upper branch and
becomes much hotter, thus allowing pressure forces in the disc to
overcome the tidal effect of the planet: the gap is closed (cf. 3D
numerical simulations by \citealt{clarkearmi03}) and the instability
front rapidly propagates through the inner disc (panel b). The typical
timescale of the evolution in this phase is the faster thermal
timescale. The three snapshots in panel b refer to $t\approx$ 35100
yr, and are separated by one year between each other. Once the inward
propagation front has reached the inner disc, the instability
continues to propagate outside much more slowly, until the maximum
extension of the propagation is reached: in the reference case this
happens at $R_{\mathrm{lim}}\approx 43R_{\odot}$. The mass accretion
rate through the inner disc is $\mdot_{\mathrm{out}} \approx 2.2~
10^{-4}\msunyr$ (panel c, snapshots separated by $\approx 5$ yrs). The
instability then slowly retreats from outside-in: in this phase the
disc is moving from the upper branch to the lower branch and is still
too hot for the planet to re-open the gap (panel d, snapshots
separated by $\approx 6$ yrs). At later times, the disc cools down to
the lower branch and the planet is now able to open the gap again
(panel e, snapshot separation $\approx 5$ yrs). The inner disc is
again drained rapidly onto he star, while the outer disc is banked up
again, being ready for another outburst to take place (panel f). This
phase now occurs on the slower viscous timescale (snapshot separation
in panel f: $\approx 1000$ yrs).

\begin{figure}
\centerline{\epsfig{figure=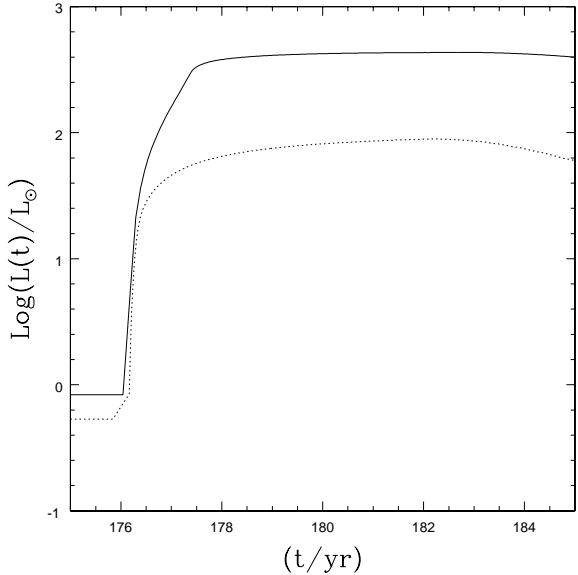,width=80mm}}
\caption{Light curves of the planet-triggered outburst (solid line),
  compared to the inside-out outburst (dotted line). We have used here
  a linear scale for the luminosity to better emphasise the
  differences between the two cases. The off-set of the temporal axis
  has been opportunely chosen for display convenience.}
\label{fig:lumref}
\end{figure}

Fig. \ref{fig:lumref} shows the light curve of the planet-induced
outburst compared to the light curve of the inside-out outburst
described in Section \ref{sec:inst}. All the parameters in these two
models are identical, except for the presence of the planet.

The first thing to notice is that the outside-in outburst triggered by
the planet indeed shows a faster rise time-scale. The maximum
luminosity is reached in $\approx 2$ yrs, whereas in the inside-out
outburst, the luminosity initially rises quickly but then slows down,
reaching the maximum in $\approx 7$ yrs.

A second important feature is that the peak luminosity is much higher
in the planet-triggered case, being $L_{\mathrm{peak}}\approx 430
L_{\odot}$ (in close agreement with the observed peak luminosity for
FU Ori), to be compared to a maximum luminosity
$L_{\mathrm{peak}}\approx 100 L_{\odot}$ for the inside-out
outburst. This can be easily understood by considering the equilibrium
solutions, equations (\ref{eq:low})-(\ref{eq:sigmaB}). In fact, during the
outburst, the inner disc accretes at an approximately constant rate,
determined roughly by the mass accretion rate in the higher branch
corresponding to the radius where the instability is first triggered,
where the surface density is $\Sigma_{\mathrm{A}}$ (i.e. to the
accretion rate denoted in Fig. \ref{fig:scurve} with
$\mu_{\mathrm{AA}}$). By combining equations (\ref{eq:high}) and
(\ref{eq:sigmaA}) we can thus obtain an estimate of the outburst mass
accretion rate:
\begin{equation}
\label{eq:mout}
\mdot_{\mathrm{out}}=3.5~10^{-5}\left(\frac{\alpha_{\mathrm{high}}}
{\alpha_{\mathrm{low}}}\right)^{16/15}
\left(\frac{R_{\mathrm{trig}}}{10R_{\odot}}\right)^{2.6}\msunyr,
\end{equation}
where $R_{\mathrm{trig}}$ is the radius at which the instability is
triggered. We then see that $\mdot_{\mathrm{out}}$ is a strongly
increasing function of the trigger radius, so that an outside-in
outburst is bound to be more violent than an inside-out one. In this
respect is very interesting to notice that, among the three best
studied FU Orionis objects, the two that show a rapid rise outburst
(i.e. FU Ori and V1057 Cyg) are indeed characterised by a larger peak
luminosity with respect to V1515 Cyg, which is characterised by a
slower rise time.

\begin{figure}
\centerline{\epsfig{figure=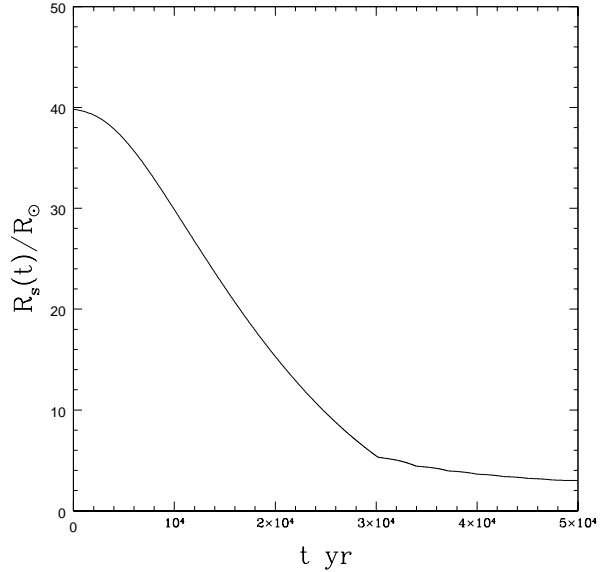,width=80mm}}
\caption{Orbital evolution of the planet. During the first $3~10^4$
  yrs the disc builds up, until the first outburst is triggered. The
  orbital evolution of the planet is significantly slowed down by the
  outburst mechanism.}
\label{fig:plan_evo}
\end{figure}

\citet{clarkesyer96} predicted that the long term evolution of the
orbit of the planet will be significantly slowed down by the outburst
mechanism. This is because the surface density left over after the
outburst will be small, and will therefore not be able to push the
planet effectively. Once the surface density has increased enough to
start moving the planet inward, then a new outburst will start and the
density will drop down again. Fig. \ref{fig:plan_evo} shows the long
term evolution of the planet position. During the first $\approx
3~10^4$ yrs the disc builds up, until the first outburst is
triggered. Subsequently the evolution of the planet is significantly
slowed down, as predicted by \citet{clarkesyer96}. Between $t\approx
3~10^4$ yrs and $t\approx 5~10^4$ yrs the system undergoes $\approx
8$ outbursts and the planet moves from $R\approx 10R_{\odot}$ to
$R\approx 3R_{\odot}$.

\begin{figure}
\centerline{\epsfig{figure=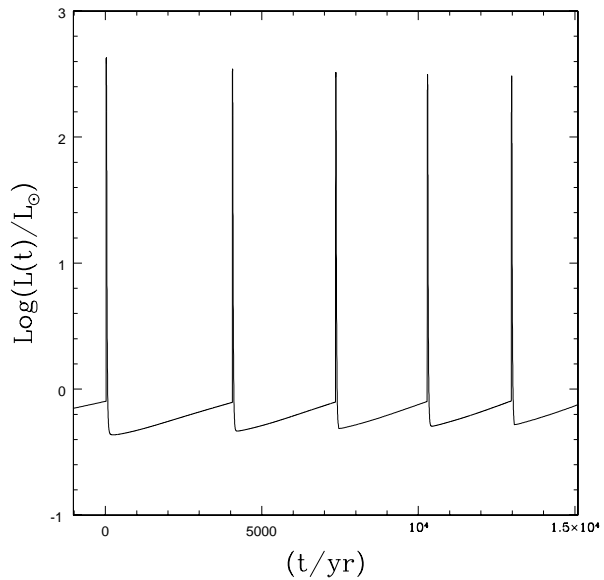,width=80mm}}
\caption{Long term evolution of the luminosity in the ``reference
  case''. The peak luminosity and the recurrence time decrease for
  late outbursts.}
\label{fig:lumlong}
\end{figure}

The slow inward migration also affects the light curves. In fact, as
the planet moves inward, the instability is subsequently triggered at
smaller radii. This causes the peak luminosity to become progressively
smaller and the recurrence time to decrease: initially, two
consecutive outbursts are separated by $\approx 4000$ yrs, while by
the end of the simulation the recurrence time has decreased to
$\approx 2000$ yr (see Fig. \ref{fig:lumlong}).

\subsection{Effect of varying parameters}
\label{sec:parameter}

\subsubsection{Planet mass}

Table \ref{tab:mass} illustrates how the properties of the outburst
depend on the mass of the planet for a fixed $\alpha$ and accretion
rate $\dot{M}_{\mathrm{in}}$. All properties are derived from the
simulations apart from $R_{\mathrm{lim,LC}}$ and
$t_{\mathrm{last,LC}}$, which are our semi-analytic predictions (see
below). Fig. \ref{fig:lum_mass} shows the corresponding light curves.

\begin{table*}
\caption{Basic outburst features with varying planet mass. See text
  for the definition of the various quantities.}
\label{tab:mass}

\begin{tabular}{@{}ccccccccc}
\hline
{$M_{\mathrm{s}}/M_{\mathrm{Jupiter}}$} &
{$\mdot_{\mathrm{out}}$ } & 
{$L_{\mathrm{peak}}/L_{\odot}$} & 
{$R_{\mathrm{trig}}/R_{\odot}$} & 
{$R_{\mathrm{lim}}/R_{\odot}$} &
{$R_{\mathrm{lim,LC}}/R_{\odot}$} & 
{$t_{\mathrm{rec}}$/yr} &
{$t_{\mathrm{last}}$/yr} &
{$t_{\mathrm{last,LC}}$/yr}\\ 
\hline 
15 & 2.5 & 500 & 11 & 46 & 47 & 4100 & 55 & 60 \\ 
10 & 2.2 & 430 & 10 & 43 & 44 & 3800 & 50 & 57 \\ 
5  & 1.6 & 300 & 8  & 38 & 40 & 3000 & 45 & 55 \\ 
2  & 1   & 190 & 6  & 32 & 34 & 2000 & 40 & 51 \\ 
\hline
\end{tabular} 

\end{table*}

\begin{figure}
\centerline{\epsfig{figure=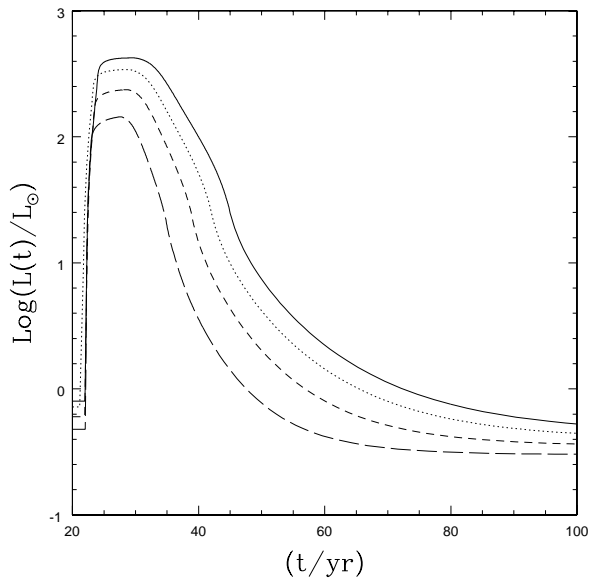,width=80mm}}
\caption{Light curves obtained by varying the planet mass. {\bf Dashed
  line}: $15M_{\mathrm{Jupiter}}$, {\bf solid line}:
  $10M_{\mathrm{Jupiter}}$ (this is the ``standard case''), {\bf long
  dashed line}: $5M_{\mathrm{Jupiter}}$, {\bf dot-dashed line}:
  $2M_{\mathrm{Jupiter}}$.}
\label{fig:lum_mass}
\end{figure}

Qualitatively, we expect $R_{\mathrm{trig}}$ to increase with planet
mass, as the disc has to bank up to higher surface densities to push
the planet inwards. In principle, we could use the analytic solutions
of \citet{ivanov99} to predict $R_{\mathrm{trig}}$ (in a manner
similar to that attempted by \citealt{clarkesyer96}), but in practice
we found that the simulation only converges to the \citet{ivanov99}
solution when the planet is about a factor 10 inside $R_{\mathrm{B}}$,
the radius at which banking up of the disc first becomes significant.
In practice, the instability is triggered when the planet is a factor of
2 or so within $R_{\mathrm{B}}$ for satellites in the planetary mass
range and so the \citet{ivanov99} solution is not applicable.

As noted already, the positive dependence of $\dot{M}_{\mathrm{out}}$
and, consequently, of $L_{\mathrm{peak}}$ on $R_{\mathrm{trig}}$ (and
hence indirectly on the planet mass) can be understood in terms of the
larger value of $\mu$ on the upper branch at $\Sigma_{\mathrm{A}}$
when $R_{\mathrm{trig}}$ is large. Equation (\ref{eq:mout}) however
overestimates $\dot{M}_{\mathrm{out}}$, since the upward transition
when the instability is triggered is not vertical in the
$\Sigma$-$\mu$ plane (see also \citetalias{bellin94}); in particular
the dependence of $\dot{M}_{\mathrm{out}}$ on $R_{\mathrm{trig}}$ is
considerably milder than that predicted by equation (\ref{eq:mout}).

The outermost location in the disc that is reached by the instability,
$R_{\mathrm{lim}}$, is an important quantity since it determines both
the outburst duration and the region of the disc contributing to the
enhanced SED during the outburst \citep{kenyon88,kenyon91}. We are
able to predict $R_{\mathrm{lim}}$ from the following simple
argument. The spike in surface density associated with the ionisation
front (see, for example, Fig. \ref{fig:reference}, panel c) must
evidently contain enough mass, when the front is at radius $R$, to
attain surface density $\Sigma_{\mathrm{A}}$. \citet{LPF85} have
argued that the front cannot be narrower than $H$, since otherwise the
radial pressure gradient would reverse the gradient of specific
angular momentum in the disc, thus rendering it Rayleigh
unstable. Thus, a lower limit to the mass in the ionisation front at
$R$ is $\Delta M(R)=2\pi R\Sigma_{\mathrm{A}}(R)H$. During the rise of
the outburst, mass diffuses out to the ionisation front from
previously destabilised regions at smaller radii. We thus derive an
estimate of the outermost propagation radius of the instability,
$R_{\mathrm{lim,LC}}$, as being the radius at which $\Delta
M(R_{\mathrm{lim,LC}})$ is equal to the mass interior to
$R_{\mathrm{lim,LC}}$ when the outburst is first triggered. Since, as
noted above, we cannot use the \citet{ivanov99} solution to predict
the surface density profile in the disc when the outburst is first
triggered, we instead obtain the interior mass from the
simulation. The resulting estimator $R_{\mathrm{lim,LC}}$ is in
excellent agreement with that obtained in the simulation. We also find
that the viscous timescale of the disc on the upper branch at
$R_{\mathrm{lim,LC}}$ provides a good estimator of outburst duration
($t_{\mathrm{last,LC}}$), if outburst duration is defined as being the
period over which the luminosity is enhanced by more than a factor 2
compared with the pre-outburst level. [We note that
\citetalias{bellin94} proposed a different estimator of
$R_{\mathrm{lim}}$, based on the minimum radius at which the disc can
remain on the lower stable branch at an accretion rate of
$\dot{M}_{\mathrm{in}}$. However, as noted by \citealt{belletal95},
$R_{\mathrm{lim}}$ can exceed this value in the case of a triggered
outburst.]

Finally, we note that the recurrence time, $t_{\mathrm{rec}}$, also
increases with planet mass, since larger planets, for which
$R_{\mathrm{lim}}$ is larger, are associated with outbursts that clear
out the disc over a larger radial range. We found, however, that the
recurrence timescale is not exactly linearly proportional to the mass
accreted onto the star during the outburst, being slightly longer than
expected for the less massive planets. We attribute this effect to the
fact that in order to replenish the disc through
$\dot{M}_{\mathrm{in}}$ during quiescence, the matter has to diffuse
to smaller radii in the case of less massive planets (since
$R_{\mathrm{trig}}$ is smaller), hence slowing down the process.

\subsubsection{Accretion rate}

\begin{figure*}
\centerline{\epsfig{figure=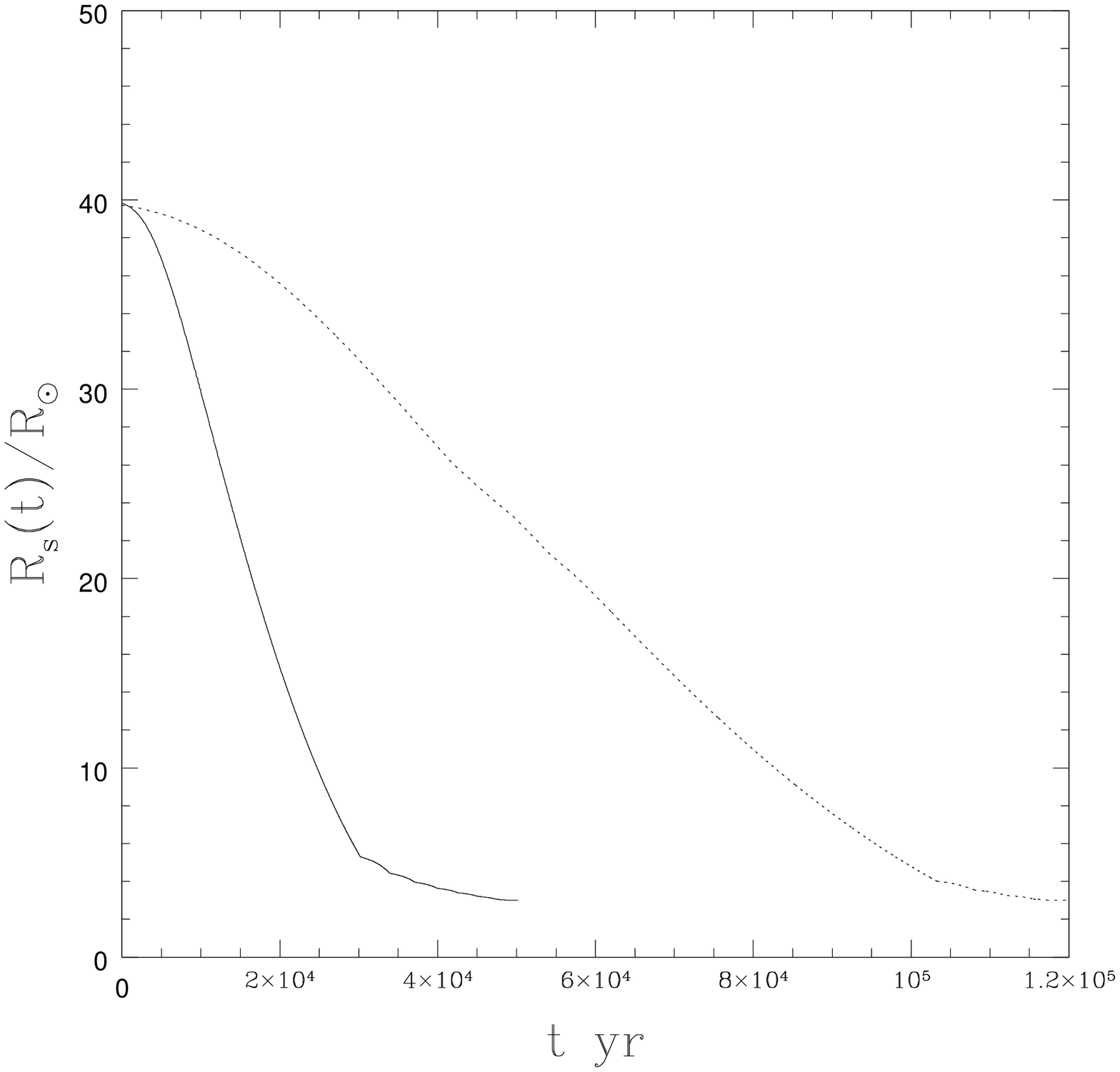,width=.5\textwidth}
            \epsfig{figure=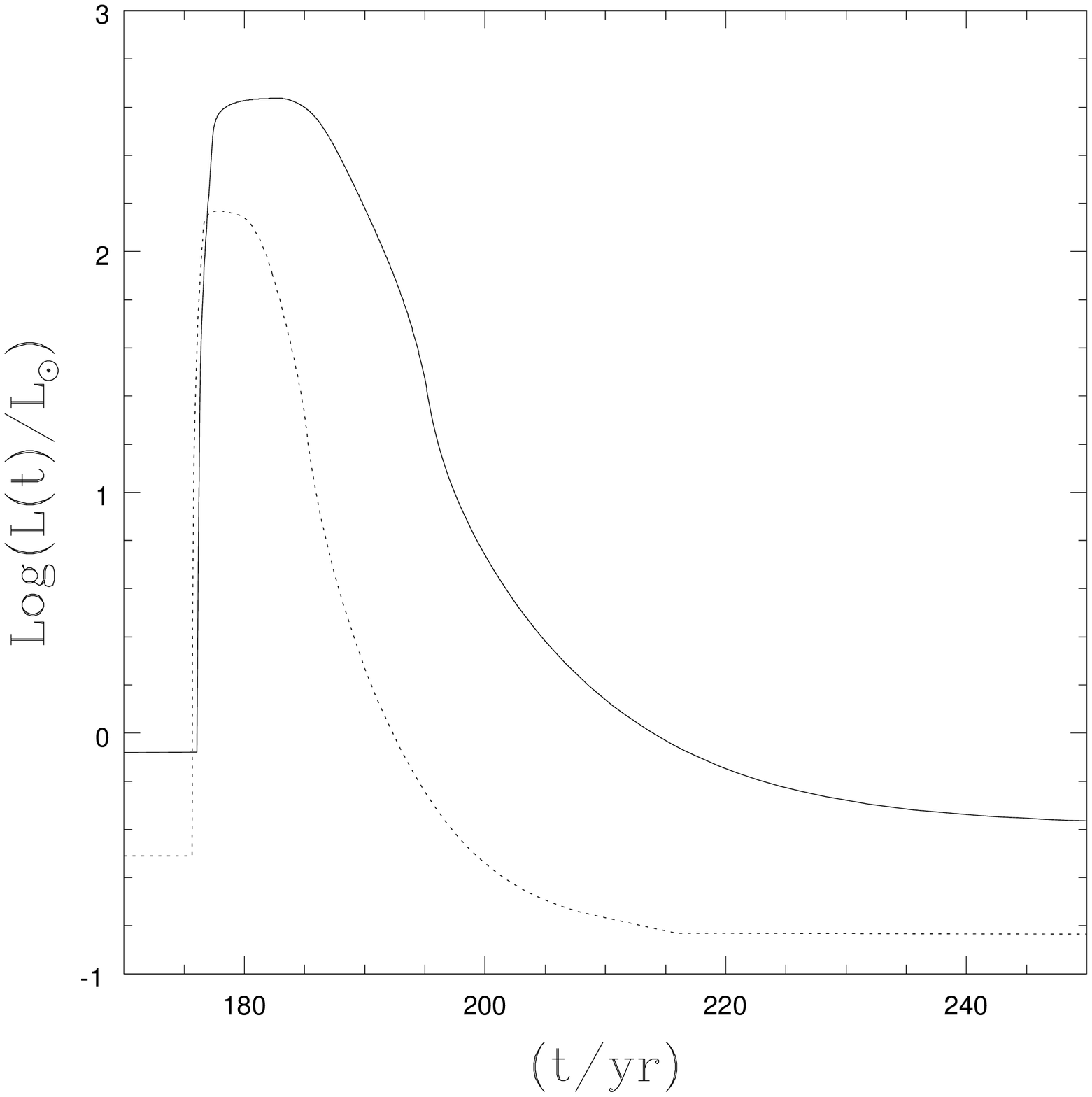,width=.5\textwidth}}
\caption{{\bf Left}: Orbital evolution of the planet for two different
  choices of $\mdot_{\mathrm{in}}= 3~10^{-6}\msunyr$ (solid line) and
  $3~10^{-7}\msunyr$ (dotted line). {\bf Right}: Corresponding light
  curves.}
\label{fig:mdotin}
\end{figure*}

When the accretion rate $\dot{M}_{\mathrm{in}}$, is lowered by an
order of magnitude to $3~10^{-7}\msunyr$ the planet spirals in more
slowly (see Fig. \ref{fig:mdotin}, left panel). This longer timescale
means that the angular momentum that the planet must lose in order to
spiral in to a given radius can be removed at a lower {\it rate}, and
hence involves a less steep banking up of the surface density in the
disc. Consequently, $R_{\mathrm{trig}}$ is attained at smaller radius
($5R_{\odot}$, compared to $10R_{\odot}$ in the reference case. The
smaller value of $R_{\mathrm{trig}}$ results in a lower amplitude,
shorter duration outburst, as expected (Fig. \ref{fig:mdotin}, right
panel).

\subsubsection{Value of $\alpha$}

\begin{figure}
\centerline{\epsfig{figure=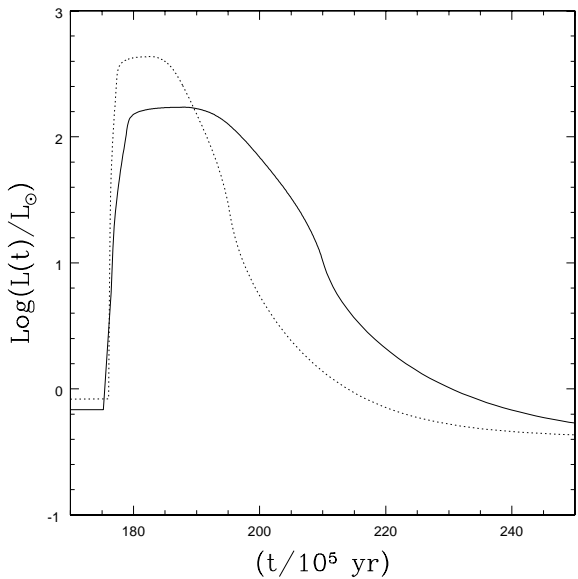,width=80mm}}
\caption{Light curves for two different choices of
  $\alpha_{\mathrm{high}}$: (solid line)
  $\alpha_{\mathrm{high}}=5~10^{-4}$, (dashed line)
  $\alpha_{\mathrm{high}}=10^{-3}$.}
\label{fig:lightalpha}
\end{figure}

The value of $\alpha$ on the high branch influences the outburst in a
manner similar to the influence of this parameter on untriggered
outbursts. Thus when $\alpha_{\mathrm{high}}$ is reduced by a factor
two, the rise and decay timescales are somewhat larger
(Fig. \ref{fig:lightalpha}). On the other hand, $\Sigma_{B}$ is
increased (Equation \ref{eq:sigmaB}) so the disc has to accrue a
smaller increase in surface density to trigger the next outburst and
consequently the recurrence time is decreased with respect to the
reference case from $\approx 3800$ years to $\approx 2000$ years.

\section{Discussion}
\label{sec:discussion}

Thermal instability models for FU Orionis objects have been
investigated previously by \citet{clarkelin90} (who investigated
outburst artificially triggered in an otherwise stable disc), by
\citetalias{bellin94} (who considered ``self-regulated'' outbursts,
produced in discs fed at a high enough rate to be thermally unstable
in the inner regions) and by \citet{belletal95} (who investigated
artificially triggered outbursts in discs fed at a high rate).

All these previous investigations have used a more refined energy
equation with respect to ours, since they have retained both the
advective and the radial radiative fluxes terms, that we neglect
here. The importance of these non-local effects has been noted
initially by \citet{clarkelin89}, in determining the outer propagation
radius of the instability. Subsequently, \citet{clarkelin90} have
noted that, in an outside-in outburst, as the instability propagates
inwards, it needs to heat up the inner disc (which is initially much
colder than the temperature of partial ionisation of hydrogen) before
making it jump to the upper branch. Therefore, initially most of the
energy gained through the enhanced accretion, rather than being
radiated away, is used to heat up the inner disc, resulting in a
sharper transition to the high luminosity phase (on the thermal
timescale of the inner disc), once the front reaches the star. In our
simulations, we are not able to follow this behaviour, since we do not
include any advective term in our energy equation. We expect that the
inclusion of these terms will result in even shorter rise times
compared to the $t_{\mathrm{rise}}\approx 2$ yrs that we find here.

The outbursts have also interesting consequences for the planet
migration. We have in fact confirmed the prediction made by
\citet{clarkesyer96} that the migration of the planet is going to be
significantly slowed down after the planet has reached the radius at
which the instability is first triggered.

If, on the one hand, we have demonstrated how the triggering effect
due to an embedded planet is able to reduce the rise timescale, which
in untriggered models would be too large compared to observations, the
duration of the outburst still remains an issue for all thermal
instability models. As we discussed above, the duration of the
outburst is related to the outermost propagation radius of the
instability, which, in thermal instability models, is bound to be
relatively small (see discussion in Sec. \ref{sec:parameter}). Even in
our longest lived outburst, $t_{\mathrm{last}}$ was not larger than
$\approx 60$ years, whereas, for example, the duration of the outburst
in FU Ori itself appears to be larger than 70 years. In order to
reproduce these long durations we needed to stretch our parameters to
rather unrealistic values, with very high $\dot{M}_{\mathrm{in}}$
(larger than $5~10^{-5}\msunyr$) and low $\alpha_{\mathrm{high}}$
($\approx 5~10^{-4}$).

Our simplified model can be refined in several ways: of course, to
obtain more realistic light curves, we should add the neglected terms
in the energy equation, in particular the advective terms, that might
influence the shape of the light curve, as discussed above. A second
limitation of our model is that we have simply assumed that the torque
between the planet and the disc is given by equation
(\ref{eq:torque}). As already discussed in Section \ref{sec:planet},
while this assumption is reasonable in the quiescent phase of the
outburst, when the planet opens up a gap in the disc, its use during
the outburst phase is questionable, since during the outburst the disc
is going to fill the gap and most of the torque exerted on the planet
might come from a circumplanetary disc. In addition, corotation
torques might also play a role in this case \citep{masset03}. Most 2D
and 3D numerical simulations of this process
\citep{lubow99,bate03,masset03} have usually considered the case of a
planet embedded in a much colder, T Tauri-like, disc and are therefore
not appropriate to test the behaviour of the planet-disc system in the
case we are interested in here. Therefore, it would be important to
perform detailed simulations of planet-disc interaction in FU Orionis
objects.

A full study of these effects is highly desirable because the early
evolution of the planet is going to be strongly influenced by the
outbursts, both because the detailed planet-disc interaction is going
to affect the planet migration and the mass accretion rate onto the
planet itself (note that in the present work the planet mass is fixed)
and because we expect that the structure of the planet will be
strongly affected by the very hot ``thermal bath'' in which the planet
is embedded during the outburst.

From a more general point of view, we note that the process described
here implies that planets are present in the early stages of star
formation, during which FU Orionis outbursts are thought to occur. On
the other hand, core accretion models of planet formation
\citep{pollack96} would imply very long timescales for the formation
of gas giant planets (this actually is one problem of the core
accretion scenario itself). Assessing the likeliness of forming a
planet early in the star formation process is beyond the scopes of
this paper and only a full theory of planet formation will give some
clues. However, we point out that {\it (i)} the planets considered
here are probably going to be swallowed by the star after a few
outbursts, and could therefore represent an early generation of
planets with only a minor fraction of them surviving the
pre-main-sequence evolution of the star; {\it (ii)} we are here
considering high mass planets, which are not easy to form via core
accretion and might have formed through some other process, such as
gravitational instabilities in the protostellar disc
\citep{boss00}. Finally, {\it (iii)}, we note that the youth of FU
Orionis objects has been generally inferred from the high input mass
accretion rates required in the \citetalias{bellin94} model to trigger
the thermal instability. Here we have shown that a planet-induced
outburst may also occur with a lower $\dot{M}_{\mathrm{in}}$, which
would allow FU Orionis outbursts to occur at a later stage in the
protostellar evolution.

\section{Conclusions}
\label{sec:conclusions}

In this paper we have shown how the interaction between a protostellar
disc and an embedded massive planet is going to naturally produce
outbursts in the disc, due to the banking up of material upstream of
the planet position, which triggers a thermal instability in the
disc. The location at which the instability is triggered depends on
the mass of the embedded planet, and is larger for more massive
planets. For a $10M_{\mathrm{Jupiter}}$ planet, we obtain
$R_{\mathrm{trig}}\approx 10R_{\odot}$. In this way, the instability
propagates outside-in, therefore producing rapid-rise outbursts (with
$t_{\mathrm{rise}}\approx 2$ yrs), as observed in some FU Orionis
objects (such as FU Ori itself and V1057 Cyg). Longer rise times, as
observed in the case of V1515 Cyg, can be obtained either in
non-triggered outbursts, of the kind discussed by
\citetalias{bellin94}, or in systems with a low-mass embedded planet
(with $M_{\mathrm{s}}\lesssim 2M_{\mathrm{Jupiter}}$).  We have also
revisited the arguments often used \citepalias{bellin94} to estimate
the outer propagation radius of the instability, $R_{\mathrm{lim}}$,
since these previous arguments are not appropriate in the case of
triggered outbursts, such as those described in this paper.

The occurrence of FU Orionis outbursts can have important effects on
the embedded planet, ({\it i}) from a dynamical point of view, since
we have demonstrated here that the outburst mechanism is going to
significantly slow down the inward migration of the planet, which is
going to linger near the position where the first outburst is
triggered (see Fig. \ref{fig:plan_evo}), and ({\it ii}) concerning the
internal structure of the planet, which as a consequence of the
outburst will be suddenly embedded in a very hot environment. We have
not discussed this last topic in the present work and we leave it to
further investigations.

\bibliographystyle{mn2e}
\bibliography{lodato}

\begin{thebibliography}{}

\bibitem[\protect\citeauthoryear{Armitage \& Bonnell}{Armitage \&
  Bonnell}{2002}]{armibonnel2002}
Armitage P.~J.,  Bonnell I.~A.,  2002, MNRAS, 330, L11

\bibitem[\protect\citeauthoryear{Armitage, Livio \& Pringle}{Armitage
  et~al.}{2001}]{armitage2001}
Armitage P.~J.,  Livio M.,    Pringle J.~E.,  2001, MNRAS, 324, 705

\bibitem[\protect\citeauthoryear{Bate, Lubow, Ogilvie \& Miller}{Bate
  et~al.}{2003}]{bate03}
Bate M.~R.,  Lubow S.~H.,  Ogilvie G.~I.,    Miller K.~A.,  2003, MNRAS, 213,
  341

\bibitem[\protect\citeauthoryear{Bell et~al.,}{Bell  et~al.}{1995}]{belletal95}
Bell K.~R.,  et~al., 1995, ApJ, 444, 376

\bibitem[\protect\citeauthoryear{Bell \& Lin}{Bell \& Lin}{1994}]{bellin94}
Bell K.~R.,  Lin D. N.~C.,  1994, ApJ, 427, 987

\bibitem[\protect\citeauthoryear{Bonnell \& Bastien}{Bonnell \&
  Bastien}{1992}]{bonnell92}
Bonnell I.,  Bastien P.,  1992, ApJ, 401, 31

\bibitem[\protect\citeauthoryear{Boss}{Boss}{2000}]{boss00}
Boss A.~P.,  2000, ApJ, 536, L101

\bibitem[\protect\citeauthoryear{Bryden, Chen, Lin, Nelson \&
  Papaloizou}{Bryden et~al.}{1999}]{bryden99}
Bryden G.,  Chen X.,  Lin D.,  Nelson R.,    Papaloizou J.,  1999, ApJ, 514,
  344

\bibitem[\protect\citeauthoryear{Bryden, R\`o\.zyczka, Lin \&
  Bodenheimer}{Bryden et~al.}{2000}]{bryden2000}
Bryden G.,  R\`o\.zyczka M.,  Lin D.,    Bodenheimer P.,  2000, ApJ, 540, 1091

\bibitem[\protect\citeauthoryear{Clarke \& Armitage}{Clarke \&
  Armitage}{2003}]{clarkearmi03}
Clarke C.~J.,  Armitage P.~J.,  2003, MNRAS, 345, 691

\bibitem[\protect\citeauthoryear{Clarke, Lin \& Papaloizou}{Clarke
  et~al.}{1989}]{clarkelin89}
Clarke C.~J.,  Lin D. N.~C.,    Papaloizou J. C.~B.,  1989, MNRAS, 236, 495

\bibitem[\protect\citeauthoryear{Clarke, Lin \& Pringle}{Clarke
  et~al.}{1990}]{clarkelin90}
Clarke C.~J.,  Lin D. N.~C.,    Pringle J.~E.,  1990, MNRAS, 242, 439

\bibitem[\protect\citeauthoryear{Clarke \& Syer}{Clarke \&
  Syer}{1996}]{clarkesyer96}
Clarke C.~J.,  Syer D.,  1996, MNRAS, 278, L23

\bibitem[\protect\citeauthoryear{Goldreich \& Tremaine}{Goldreich \&
  Tremaine}{1980}]{goldreich80}
Goldreich P.,  Tremaine S.,  1980, ApJ, 241, 425

\bibitem[\protect\citeauthoryear{Hartmann \& Kenyon}{Hartmann \&
  Kenyon}{1996}]{hartmann96}
Hartmann L.,  Kenyon S.~J.,  1996, ARA\&A, 34, 207

\bibitem[\protect\citeauthoryear{Herbig}{Herbig}{1977}]{herbig77}
Herbig G.~H.,  1977, ApJ, 217, 693

\bibitem[\protect\citeauthoryear{Herbig, Petrov \& Duemmler}{Herbig
  et~al.}{2003}]{herbig03}
Herbig G.~H.,  Petrov P.~P.,    Duemmler R.,  2003, ApJ, 595, 384

\bibitem[\protect\citeauthoryear{Ivanov, Papaloizou \& Polnarev}{Ivanov
  et~al.}{1999}]{ivanov99}
Ivanov P.~B.,  Papaloizou J. C.~B.,    Polnarev A.~G.,  1999, MNRAS, 307, 79

\bibitem[\protect\citeauthoryear{Kenyon \& Hartmann}{Kenyon \&
  Hartmann}{1991}]{kenyon91}
Kenyon S.~J.,  Hartmann L.,  1991, ApJ, 383, 664

\bibitem[\protect\citeauthoryear{Kenyon, Hartmann \& Hewett}{Kenyon
  et~al.}{1988}]{kenyon88}
Kenyon S.~J.,  Hartmann L.,    Hewett R.,  1988, ApJ, 325, 231

\bibitem[\protect\citeauthoryear{Lasota}{Lasota}{2001}]{lasota01}
Lasota J.~P.,  2001, New Astronomy Reviews, 45, 449

\bibitem[\protect\citeauthoryear{Lin \& Papaloizou}{Lin \&
  Papaloizou}{1979a}]{linpap79b}
Lin D.,  Papaloizou J.,  1979a, MNRAS, 188, 191

\bibitem[\protect\citeauthoryear{Lin \& Papaloizou}{Lin \&
  Papaloizou}{1986}]{linpap86b}
Lin D.,  Papaloizou J.,  1986, ApJ, 309, 846

\bibitem[\protect\citeauthoryear{Lin, Papaloizou \& Faulkner}{Lin
  et~al.}{1985}]{LPF85}
Lin D.,  Papaloizou J.,    Faulkner J.,  1985, MNRAS, 212, 105

\bibitem[\protect\citeauthoryear{Lin \& Papaloizou}{Lin \&
  Papaloizou}{1979b}]{linpap79a}
Lin D. N.~C.,  Papaloizou J.,  1979b, MNRAS, 186, 799

\bibitem[\protect\citeauthoryear{Lodato \& Bertin}{Lodato \&
  Bertin}{2001}]{LB2001}
Lodato G.,  Bertin G.,  2001, A\&A, 375, 455

\bibitem[\protect\citeauthoryear{Lodato \& Bertin}{Lodato \&
  Bertin}{2003}]{LB03b}
Lodato G.,  Bertin G.,  2003, A\&A, 408, 1015

\bibitem[\protect\citeauthoryear{Lubow, Seibert \& Artymowicz}{Lubow
  et~al.}{1999}]{lubow99}
Lubow S.~H.,  Seibert M.,    Artymowicz P.,  1999, ApJ, 526, 1001

\bibitem[\protect\citeauthoryear{Masset \& Papaloizou}{Masset \&
  Papaloizou}{2003}]{masset03}
Masset F.,  Papaloizou J. C.~B.,  2003, MNRAS, 588, 494

\bibitem[\protect\citeauthoryear{Meyer \& Meyer-Hofmeister}{Meyer \&
  Meyer-Hofmeister}{1981}]{meyer81}
Meyer F.,  Meyer-Hofmeister E.,  1981, A\&A, 104, 10

\bibitem[\protect\citeauthoryear{Natarajan \& Armitage}{Natarajan \&
  Armitage}{2002}]{natarajan02}
Natarajan P.,  Armitage P.~J.,  2002, ApJ, 567, L9

\bibitem[\protect\citeauthoryear{Pollack et~al.,}{Pollack
  et~al.}{1996}]{pollack96}
Pollack J.~B.,  et~al., 1996, Icarus, 124, 62

\bibitem[\protect\citeauthoryear{Potter}{Potter}{1973}]{potter}
Potter D.,  1973, Computational Physics.
Wiley and Sons, London

\bibitem[\protect\citeauthoryear{Pringle}{Pringle}{1981}]{pringle81}
Pringle J.~E.,  1981, ARA\&A, 19, 137

\bibitem[\protect\citeauthoryear{Pringle, Verbunt \& Wade}{Pringle
  et~al.}{1986}]{pringle86}
Pringle J.~E.,  Verbunt F.,    Wade R.~A.,  1986, MNRAS, 221, 169

\bibitem[\protect\citeauthoryear{Shakura \& Sunyaev}{Shakura \&
  Sunyaev}{1973}]{shakura73}
Shakura N.~I.,  Sunyaev R.~A.,  1973, A\&A, 24, 337

\bibitem[\protect\citeauthoryear{Syer \& Clarke}{Syer \&
  Clarke}{1995}]{clarkesyer95}
Syer D.,  Clarke C.~J.,  1995, MNRAS, 277, 758

\bibitem[\protect\citeauthoryear{Takeuchi, Miyama \& Lin}{Takeuchi
  et~al.}{1996}]{takeuchi96}
Takeuchi T.,  Miyama S.~M.,    Lin D. N.~C.,  1996, ApJ, 460, 832

\bibitem[\protect\citeauthoryear{Tanaka, Takeuchi \& Ward}{Tanaka
  et~al.}{2002}]{tanaka02}
Tanaka H.,  Takeuchi T.,    Ward W.~R.,  2002, ApJ, 565, 1257

\bibitem[\protect\citeauthoryear{Trilling, Benz, Guillot, Lunine, Hubbard \&
  Burrows}{Trilling et~al.}{1998}]{trilling98}
Trilling D.,  Benz W.,  Guillot T.,  Lunine J.~I.,  Hubbard W.~B.,    Burrows
  A.,  1998, ApJ, 500, 428

\end{thebibliography}

\end{document}